\documentclass[final]{elsarticle}
\usepackage{amssymb}

\usepackage{graphicx}
\usepackage{amsmath,bm}
\usepackage{amsfonts}
\usepackage[justification=centering]{caption}
\usepackage[font={small}]{caption}
\usepackage{epstopdf}
\usepackage{url}
\usepackage{array}
\usepackage{textcomp}
\newcolumntype{C}[1]{>{\centering\arraybackslash$}p{#1}<{$}}
\usepackage{enumerate}
\usepackage{subcaption} 

\usepackage{algorithm} 
\usepackage{algpseudocode}
\usepackage{pifont}

\usepackage{epstopdf}
\usepackage{xcolor}

\journal{Neurocomputing}

\begin{document}
    \makeatletter
    \def\ps@pprintTitle{%
       \let\@oddhead\@empty
       \let\@evenhead\@empty
       \let\@oddfoot\@empty
       \let\@evenfoot\@oddfoot
    }
    \makeatother
    
\begin{frontmatter}

\title{Double Temporal Sparsity Based Accelerated
Reconstruction in Compressed Sensing fMRI}

\author[rvt]{Priya Aggarwal\corref{cor1}}
\ead{priyaa@iiitd.ac.in}
\author[rvt]{Anubha Gupta}
\ead{anubha@iiitd.ac.in}
\cortext[cor1]{Corresponding author}
\address[rvt]{Signal processing and Bio-medical Imaging Lab, Department of Electronics and Communication Engineering, Indraprastha Institute of Information Technology (IIIT), Delhi, India}


\begin{abstract}
A number of reconstruction methods have been proposed recently for accelerated functional Magnetic Resonance Imaging (fMRI) data collection. However, existing methods suffer with the challenge of greater artifacts at high acceleration factors. This paper addresses the issue of accelerating fMRI collection via undersampled \textit{k}-space measurements combined with the proposed Double Temporal Sparsity based Reconstruction (DTSR) method with the $l^{1}-l^{1}$ norm constraint. The robustness of the proposed DTSR method has been thoroughly evaluated both at the subject level and at the group level on real fMRI data. Results are presented at various acceleration factors. Quantitative analysis in terms of Peak Signal-to-Noise Ratio (PSNR) and other metrics, and qualitative analysis in terms of reproducibility of brain Resting State Networks (RSNs) demonstrate that the proposed method is accurate and robust. In addition, the proposed DTSR method preserves brain networks that are important for studying fMRI data. Compared to the existing accelerated fMRI reconstruction methods, the DTSR method shows promising potential with an improvement of 10-12dB in PSNR with acceleration factors upto 3.5. Simulation results on real data demonstrate that DTSR method can be used to acquire accelerated fMRI with accurate detection of RSNs.
\end{abstract}

\end{frontmatter}

\section{Introduction}
Functional Magnetic Resonance Imaging (fMRI) is a prominent and widely used noninvasive neuroimaging modality \cite{fMRIref1,Feinberg2012, Ogawa1990}. It is used to understand brain functions by measuring Blood Oxygen Level Dependent (BOLD) signals. BOLD  fMRI signals are captured via T$2^{*}$  weighted imaging. Achieving high temporal resolution remains challenging in fMRI. In addition, one of the major challenges in fMRI is the low sensitivity of BOLD signals leading to images blurred with motion and other artifacts. Poor sensitivity of BOLD signals leads to lower Signal-to-Noise Ratio (SNR). Moreover, long scanning times lead to annoyance in patients. Hence, there is a need to capture images in the shortest possible time. 

Various remedies have been proposed to address high spatio-temporal resolution of fMRI such as development of high magnetic field scanner \cite{Duong2002, Yacoub2003, Harel2012,Yacoub2008}, coil sensitivity improvement inside fMRI scanner \cite{Logothetis2002}, advancements in pulse sequences \cite{Pfeuffer2002,Wu2013}, usage of parallel imaging \cite{Hugger2011, Chaari2014}, and compressed sensing (CS) based fMRI reconstruction from fewer $k$-space (spatial Fourier domain) measurements \cite{modCS,Yan2014,han2015,Jung2009,Zong2014,chavarrias2015,Holland2013,Fang2015,Chiew2016,Singh2015}. In this paper, we address the problem of accelerated fMRI reconstruction without the loss of BOLD sensitivity in the CS framework.

Compressive sampling involves capturing of less data \cite{Donoho2006}. Since acquisition time is directly related to the number of measurements, CS reduces the data acquisition time in fMRI. CS allows reconstruction of fMRI brain volumes using less number of $\textit{k}$-space measurements that are picked up at a sampling rate lower than the Nyquist sampling rate, provided some assumptions hold true. In general, it is assumed that the data is sparse in some transform domain and that the chosen samples are incoherent \cite{han2015}. Compressive sensing framework helps in fMRI reconstruction in two significant ways: 1) It helps in increasing the statistical power of the BOLD signal \cite{Jung2009,Holland2013} because of its inherent denoising property and 2) it provides improvement in the spatiotemporal resolution of fMRI data \cite{Zong2014,Chiew2016,modCS,Singh2015,Yan2014,chavarrias2015,Fang2015,han2015,Aggarwal2017,Aggarwal2016}.

Various CS based reconstruction methods have been proposed so far for accelerated fMRI reconstruction \cite{modCS,Yan2014,han2015,Jung2009,Zong2014,chavarrias2015,Holland2013,Fang2015,Chiew2016,Singh2015,Aggarwal2016,Aggarwal2017}. 
These methods can be largely divided into two categories. First category includes online methods that reconstruct fMRI data in real time \cite{modCS,Yan2014,han2015,Jung2009,Zong2014}. These methods reconstruct brain volume at time \textit{t} using the volume at time $t-1$ by assuming causality in the reconstruction framework. Second category includes offline methods that first store \textit{k}-space data of all fMRI volumes and later, utilize this complete information across both time and space, also called $\textit{k-t}$ space data, to reconstruct fMRI data \cite{chavarrias2015,Holland2013,Fang2015,Chiew2016,Singh2015,Aggarwal2016,Aggarwal2017}.

Many offline reconstruction methods, such as Compressed sensing with wavelet domain sparsity (CSWD) \cite{Holland2013}, HSPARSE \cite{Fang2015}, \textit{k-t} FASTER \cite{Chiew2016}, and LR+S \cite{Singh2015}, have been proposed in the fMRI literature. These methods largely differ in regularization constraints in CS fMRI reconstruction framework. CS fMRI solves a set of underdetermined equations that has infinitely many solutions. In order to recover a unique solution corresponding to the signal of interest, regularization constraints are added. Often, sparsity in some transform domain is added as the regularization constraint. Theoretical studies have shown that it is possible to recover sparse signals by $\textit{l}^{1}$ norm minimization \cite{Candes2006}. 

For example, in \cite{Holland2013}, undersampled fMRI data is reconstructed using compressed sensing with sparsity of fMRI data in the wavelet domain, wherein orthogonal Daubechies wavelet is used as the sparsifying basis. In \cite{Holland2013}, fMRI data is reconstructed using $l^{1}$ norm constraint by assuming sparsity in the temporal direction. Recently, in \cite{Fang2015}, both temporal and spatial sparsity are exploited to recover fMRI data. Here, CS is also utilized to gain high spatial resolution fMRI and method is named as High Spatial Resolution Compressed Sensing (HSPARSE) \cite{Fang2015}. 

In \cite{Chiew2016}, \textit{k-t} FASTER method is proposed that recovers a low rank signal via iterative hard thresholding of singular values of data matrix in the CS framework. In another work \cite{Singh2015,Aggarwal2017}, fMRI reconstruction is performed using low-rank plus sparse (LR+S) decomposition of fMRI signals. Here, an iterative framework is used that reconstructs the low rank and the sparse components of fMRI data separately.

In this paper, we introduce a novel offline fMRI reconstruction method. In fMRI, same brain volumes are scanned repeatedly over time in order to study brain's function. This fact is used as an advantage in the proposed method  via total variation based regularization \cite{wang2008} because scanning of the same brain volume over time brings similarity in the temporal direction. In addition, we impose conventional temporal sparsity in the proposed reconstruction framework and hence, name the proposed method as Double Temporal Sparsity based Reconstruction (DTSR).

We thoroughly evaluate the robustness and the feasibility of the proposed DTSR method both at subject and group levels of real fMRI data analysis. The performance of DTSR method has been preliminary evaluated using retrospective undersampling of the fully available fMRI dataset. We compare the performance of the proposed method with other offline methods such as Compressed Sensing with Wavelet domain Sparsity (CSWD) \cite{Holland2013}, HSPARSE \cite{Fang2015}, \textit{k-t} FASTER \cite{Chiew2016}, and LR+S \cite{Singh2015}. Results demonstrate that DTSR is able to improve BOLD sensitivity both at the individual and at the group level compared to other methods. Existing reconstruction methods produce greater artifact at high acceleration factors. The proposed DTSR reconstruction method increases temporal resolution without affecting the spatial resolution and can be used to provide accelerated high temporal resolution fMRI reconstruction with accurate detection of intrinsic brain's Resting State Networks (RSNs). \\

Key contributions of this work are summarized as below:
\begin{itemize}
\item A double temporal-sparsity based reconstruction framework is proposed for the robust reconstruction of undersampled fMRI data.

\item An algorithm is designed to solve the proposed DTSR reconstruction approach using the state-of-the-art Alternating Direction Multiplier Method (ADMM). 

\item The performance of our proposed method is
evaluated on real dataset using both quantitative and qualitative analysis.  Quantitative analysis in terms of Peak Signal-to-Noise Ratio (PSNR) and other metrics, and qualitative analysis in terms of reproducibility of brain RSNs demonstrate the robustness and efficacy of the proposed DTSR reconstruction method.
\end{itemize}

This paper is organized into five sections. In the Materials and Methods section, we discuss dataset description and some preliminary theory. In this section, we also present the proposed DTSR reconstruction method. We present experimental results on fMRI data in Section 3. We provide a thorough discussion of reconstruction results in section 4. In the end, conclusions are presented in section 5.

\section{Materials and Methods}

\subsection{Dataset Description}
The proposed DTSR method has been evaluated on the freely available Beijing\_Zang  resting state fMRI dataset. This real fMRI dataset is a part of Neuroimaging Informatics Tools and Resources Clearinghouse (NITRC) 1000 functional connectomes project \cite{biswal2010}. 

It consists of an acquisition of 33 axial interleaved ascending brain slices with a dimension of 64x64 at each time point with Repetition Time (TR) equal to 2 seconds. The fMRI brain data is collected over 225 time points. This dataset consists of 198 subjects' resting-state fMRI data, aged between 18 to 26 years of age and acquired while subjects' eyes were closed. For more details on this dataset, please refer to the website\footnote{\url{http://fcon_1000.projects.nitrc.org/}}.
For this paper, data of first 20 subjects from the set of 198 Beijing\_Zang subjects has been downloaded from the 1000 functional connectomes project's online database.

\subsection{Preliminaries}

This subsection presents the CS-based fMRI reconstruction problem formulation. Consider a 4-D fMRI dataset $\mathbb{R}^{n_{x}\times n_{y}\times n_{z}\times T}$, where $n_{z}$ denotes the number of brain slices with each slice of dimension $n_{x}\times n_{y}$ and $T$ denotes the number of time points. In fMRI, 3-D brain volumes (each volume is made up of $n_{z}$ number of brain slices) are captured repeatedly over $T$ number of time points. 

Consider a Casorati matrix $\textbf{X}=[\textbf{x}_{1},\textbf{x}_{2},...,\textbf{x}_{T}]\in \mathbb{R}^{n\times T}$ of any brain slice, where $n=n_{x}\times n_{y}$ are the number of voxels in that brain slice and $T$ denotes the total number of time points. Each column $\textbf{x}_{t}\in \mathbb{R}^{n}$ of $\textbf{X}$ represents the brain slice at the $t^{th}$ time point. In accelerated fMRI, less amount of slice data is captured in the $k$-space in order to achieve quicker scanning of slices. This allows capturing of more brain volumes in a given time leading to higher temporal resolution. Consider $\textbf{Y}$ to be such a compressively sensed $k$-space data that can be represented as 

\begin{equation}
\mathbf{Y}=\mathbf{\Phi} \mathbf{FX}+ \bm{\xi},
\label{eq:no1DTSR}
\end{equation}
where $\textbf{F}$ denotes the 2-D Fourier transform operator applied on the Casorati matrix $\textbf{X}$ for transforming this data to the $k$-space domain, $\mathbf{\Phi}$ is the sensing matrix that contains information about the partial measurements in the $k$-space domain, and $\bm{\xi} \in \mathbb{R}^{n\times T}$ denotes the measurement noise. Given a sensing matrix $\mathbf{\Phi}$, the aim of any fMRI reconstruction method is to recover Casorati data matrix $\textbf{X}$ from partial Fourier measurements $\textbf{Y}$. Reconstruction is done independently for all $n_{z}$ brain slices.

The task of computing \textbf{X} from \textbf{Y} is an underdetermined inverse problem. It is not possible to solve \eqref{eq:no1DTSR} by computing the inverse because the sensing matrix $\mathbf{\Phi}$ in the forward equation is usually ill-conditioned due to large undersampling. Therefore, in general the problem needs to be regularized to find a solution. A relatively simple solution is a well known sparsity regularization \cite{Holland2013}.  

Sparse recovery  methods assume the desired signal to be sparse over some known \textit{apriori} transform basis $\mathbf{\Psi}$ and hence, $l^{1}$ norm in the corresponding domain is used as regularization to recover the signal. $l^{1}$ norm is used as a surrogate for standard sparsity inducing $l^{0}$ norm because regularization over $l^{0}$ norm is a non-deterministic polynomial (NP) hard problem. Thus, using sparsity regularization, fMRI reconstruction problem can be formulated as

\begin{equation}
\hat{\mathbf{X}}=arg ~\underset{\mathbf{X}}{min} \left \| \mathbf{Y}-\mathbf{\Phi} \mathbf{FX} \right \|_{F}^{2}+\lambda_{1}\left \| \mathbf{\Psi}\mathbf{X} \right \|_{1 },
\label{eq:no2DTSR}
\end{equation}	
where $\left \| \cdot  \right \|_{F}^{2}$ denotes the Frobenius norm that is defined as $\left \| \mathbf{Y}-\mathbf{\Phi} \mathbf{FX} \right \|_{F}^{2}=Tr[(\mathbf{Y}-\mathbf{\Phi} \mathbf{FX})^{T}(\mathbf{Y}-\mathbf{\Phi} \mathbf{FX})]$, $\mathbf{\Psi}$ denotes the sparsifying transform basis, $\left \| \cdot  \right \|_{1}$ is the $l^{1}$ norm, and $\lambda_{1}$ is the regularization parameter that governs the sparsity on $\textbf{X}$ over the $\bm{\Psi}$ basis. The first term in \eqref{eq:no2DTSR} is the data fidelity term that minimizes the variance of noise $\mathbf{\xi}$, while the second term is the sparsity promoting term. In general, Forbenius norm and $l^{1}$ norm of any matrix $\textbf{A}\in \mathbb{R}^{n\times T}$ are defined as

\begin{equation}
\left \| \mathbf{A} \right \|_{F}^{2}=\sum_{i=1}^{n}\sum_{j=1}^{T}a_{ij}^{2}~~~~ \text{and} ~~~~
\left \| \mathbf{A} \right \|_{1}=\sum_{i=1}^{n}\sum_{j=1}^{T}\left | a_{ij} \right |.
\label{eq:no3}
\end{equation}

\subsection{Proposed DTSR Method}
\begin{figure*}[!ht]
\begin{center}
\begin{subfigure}[b]{\textwidth}
\includegraphics[width=1.1\textwidth, height=0.15\textheight, trim =40mm 1mm 1mm 1mm]{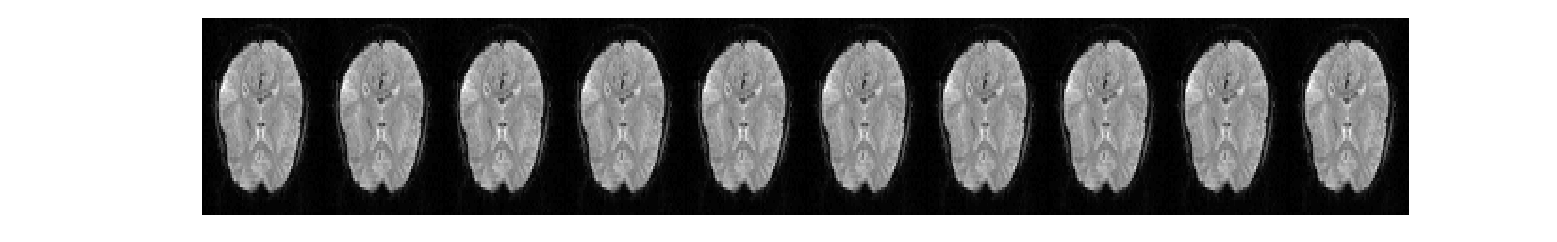}
 \caption{Middle slice at 10 consecutive time points of one subject}
\end{subfigure}

\begin{subfigure}[b]{\textwidth}
\includegraphics[width=1.12\textwidth, height=0.15\textheight, trim =40mm 1mm 1mm 0mm]{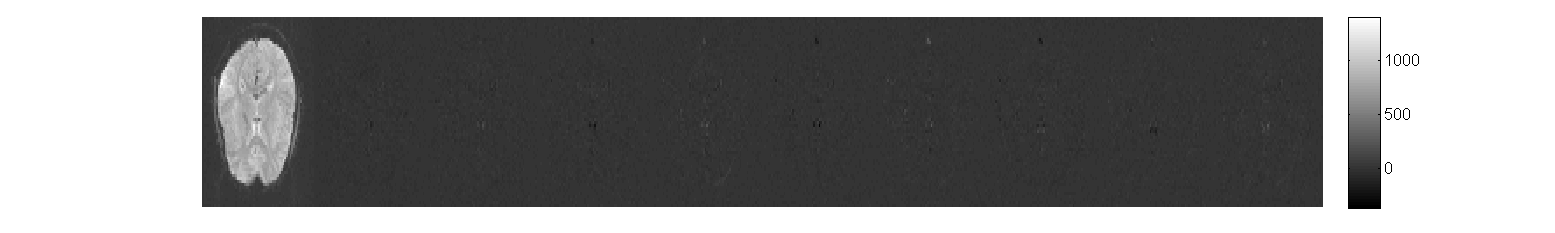}
\caption{Difference images to show total variation temporal sparsity, where difference is computed over successive slices at \textit{t} and \textit{t}-1 time points. First time point of brain slice is considered without differencing as shown.}\end{subfigure}
\end{center}
\caption{Illustration of total variation temporal sparsity on middle slice of one subject's fMRI data. Total variation temporal sparsity method arises due to repeated scanning of the same brain volume in fMRI to study brain's function. Scanning of the same brain volume over time brings similarity in the temporal direction that can be utilized via total variation based regularization. }\label{figure:no1DTSR}
\vspace{-2mm}
\end{figure*}

Consider the Casorati matrix \textbf{X} for any brain slice, where each column $\textbf{x}_{t}$ represents the vectorized brain slice captured at the $t^{th}$ time point. Since a brain slice over adjacent time points may contain less amplitude changes, the difference of adjacent columns of this Casorati matrix exhibits strong sparsity. Fig.\ref{figure:no1DTSR} illustrates sparsity of the first difference between two consecutive axial brain slices. From this figure, it is reasonable to assume that the successive difference of columns of Casorati matrix is sparse, i.e., the amplitude difference of slice at time point \textit{t} and \textit{t}-1 is sparse. This form of difference sparsity is also known as total variation in the case of 1-D signal recovery \cite{wang2008}. We call this sparsity in the context of fMRI as total variation temporal sparsity because it exploits sparsity in the temporal direction. In addition to this, we also impose conventional temporal sparsity, i.e., the sparsity of the temporal data in some transform domain and hence, name the proposed method as Double Temporal Sparsity based Reconstruction (DTSR). 

In the total variation temporal sparsity, the difference matrix of $\textbf{X}$ is assumed to be sparse. This difference matrix is formed by performing the first difference on the consecutive columns of $\textbf{X}$. First differencing is performed from $2^{\text{nd}}$ column onwards. Thus, the DTSR objective function can be formulated as

\begin{equation}
\begin{split}
\hat{\mathbf{X}}=arg ~\underset{\mathbf{X}}{min} \left \| \mathbf{Y}-\mathbf{\Phi} \mathbf{FX} \right \|_{F}^{2}+\lambda_{1}\left \| \mathbf{\Psi}\mathbf{X} \right \|_{1 }+\\
\lambda_{2}\sum_{t=2}^{T}\left | \mathbf{x}_{t}-\mathbf{x}_{t-1} \right |,
\end{split}
\label{eq:no4DTSR}
\end{equation}
where $\lambda_{1}$ and $\lambda_{2}$ are non-negative regularization parameters. Since second regularization term in \eqref{eq:no4DTSR} is non-differentiable, it is not easy to solve DTSR in this formulation. Thus, \eqref{eq:no4DTSR} is reformulated below with matrix version that provides efficient solution to this problem.
 
\begin{equation}
\hat{\mathbf{X}}=arg ~\underset{\mathbf{X}}{min} \left \| \mathbf{Y}-\mathbf{\Phi} \mathbf{FX} \right \|_{F}^{2}+\lambda_{2}\left \| \mathbf{\Psi}\mathbf{X} \right \|_{1 }+\lambda_{3}\left \| \mathbf{X}\mathbf{D} \right \|_{1 }.
\label{eq:no5DTSR}
\end{equation}	

Here, D performs first order differencing on the successive columns of the given matrix \textbf{X} and is defined as:

\begin{equation*}
\textbf{D}=\begin{bmatrix}
\begin{matrix}
-1 & 1 & 0 &.  & . & 0 &0 \\ 
 0& -1 & 1 & 0 & . & . &0 \\ 
 0& 0 & -1 &1  & 0 & . &. \\ 
 0& 0 & . &  .& . & . & .\\ 
 .& . &  .& . & . & . & 0\\ 
 0&.  & . & . &  0& -1 & 1\\ 
 0& 0 & . & 0 &0  & 0 & -1
 \end{matrix}
\end{bmatrix}.
\label{D}
\end{equation*} 

\begin{figure*}[!ht]
  \begin{subfigure}[b]{1\columnwidth}
  \centering
    \includegraphics[scale=0.8, trim = -10 0 0 0]{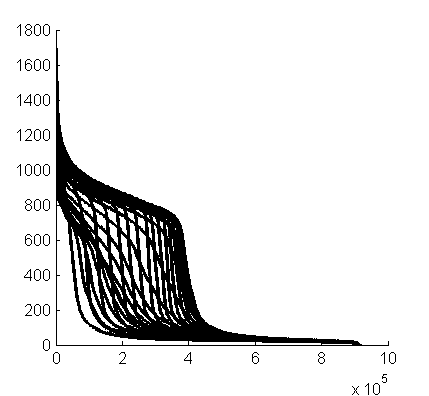}
    \caption{Decay of coefficients of $\textbf{X}$}
  \end{subfigure}
  \begin{subfigure}[b]{\columnwidth}
  \centering
    \includegraphics[scale=0.8, trim = -10 0 0 0]{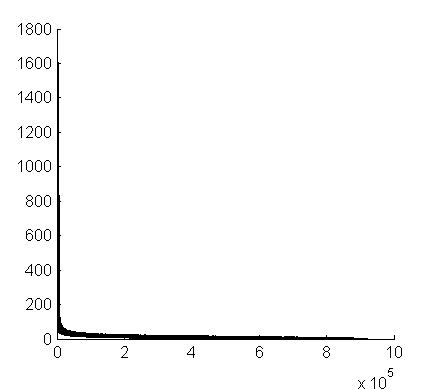}
    \caption{Decay of coefficients of $\textbf{X}\textbf{D}$}
  \end{subfigure}
\caption{Illustration of sparsity of \textbf{X} and \textbf{XD}, where $\textbf{X}$ represents a brain slice stacked over all time points and D performs first differencing on the successive columns of the given matrix \textbf{X}, or in other words performs total variation temporal sparsity. Fig.\ref{figure:no2DTSR}a and Fig.\ref{figure:no2DTSR}b plot sorted coefficients of \textbf{X} and \textbf{XD} corresponding to a real dataset. All 33 slices of one randomly selected subject's data is used in these figures. This figure indicates that the sparsity assumption on \textbf{XD} is reasonably valid because Fig.\ref{figure:no2DTSR}b has more sparse coefficients compared to Fig.\ref{figure:no2DTSR}a.} \label{figure:no2DTSR}
\end{figure*}

The total variation temporal sparsity is illustrated on real dataset in Fig.\ref{figure:no2DTSR}a and Fig.\ref{figure:no2DTSR}b that plot sorted values of \textbf{X} and \textbf{XD}, respectively. All 33 slices of one randomly selected subject's data is used in these figures. This figure indicates that the sparsity assumption on \textbf{XD} is reasonably valid.

\subsubsection{Algorithm Design}
In this subsection, an algorithm is designed to solve \eqref{eq:no5DTSR} using the alternating direction multiplier method (ADMM) \cite{boyd2011}. ADMM is suitable for constrained optimization problems and is being used extensively since past few years \cite{Chen2016,Monti2014,Zheng2013,lingala2011}. This technique facilitates solution by decomposing the original objective function into multiple objective functions that are easy to solve.

Following \cite{boyd2011}, two auxiliary matrices $\textbf{W}\in \mathbb{R}^{n\times T}$ and $\textbf{Z}\in \mathbb{R}^{n\times T}$ are introduced in \eqref{eq:no5DTSR} as

\begin{equation}
\begin{split}
\hat{\mathbf{X}}=arg ~\underset{\mathbf{X}}{min} \left \| \mathbf{Y}-\mathbf{\Phi} \mathbf{FX} \right \|_{F}^{2}+\lambda_{1}\left \| \mathbf{W} \right \|_{1 }++\lambda_{2}\left \| \mathbf{Z} \right \|_{1 } \\~~~~~s.t. ~~\textbf{W}=\mathbf{\Psi}\textbf{X},~ \textbf{Z}=\textbf{XD}.
\label{eq:no6DTSR}
\end{split}
\end{equation}	

In addition to this, constraints with equality are added for each of the auxiliary matrices. Thus, the new objective function is written as:
\begin{equation}
\begin{split}
arg \underset{\mathbf{X,Z,W}}{min} \left \| \mathbf{Y}-\mathbf{\Phi} \mathbf{FX} \right \|_{F}^{2}+\lambda _{1}\left \| \mathbf{W} \right \|_{1 }+\lambda _{2}\left \|  \mathbf{Z} \right \|_{1}+\\
\frac{\eta _{1}}{2}\left \| \mathbf{W-\mathbf{\Psi}X-B_{1}} \right \|_{F}^{2}+\frac{\eta _{2}}{2}\left \| \mathbf{Z-XD-B_{2}} \right \|_{F}^{2},
\label{eq:no7DTSR}
\end{split}
\end{equation}
where $\eta _{1}$ and $\eta _{2}$ are penalty parameters and, $\textbf{B}_{1}$ and $\textbf{B}_{2}$ are the Lagrange multipliers used to enforce equality between the original and auxiliary matrices. 

ADMM updates variables \textbf{W}, \textbf{Z}, and \textbf{X} alternately in the above defined augmented Lagrangian function. The minimization over one variable in an iteration assumes the other two variables to be fixed. Therefore, the above function can be alternately optimized over each variable separately. This allows splitting of \eqref{eq:no7DTSR} into different subproblems with three new objective functions stated as below:

\begin{equation}
\begin{split}
P1: ~~arg ~\underset{\mathbf{W}}{min} ~~\lambda _{1}\left \| \mathbf{W} \right \|_{1 }+ \frac{\eta _{1}}{2}\left \| \mathbf{W-\mathbf{\Psi}X^{\textit{j-1}}-B_{1}^{\textit{j-1}}} \right \|_{F}^{2},
\label{eq:no8DTSR}
\end{split}
\end{equation}

\begin{equation}
\begin{split}
P2: ~~arg~ \underset{\mathbf{Z}}{min} ~~\lambda _{2}\left \|  \mathbf{Z} \right \|_{1}+\frac{\eta _{2}}{2}\left \| \mathbf{Z-X^{\textit{j-1}}D-B_{2}^{\textit{j-1}}} \right \|_{F}^{2},
\label{eq:no9DTSR}
\end{split}
\end{equation}

\begin{equation}
\begin{split}
P3: ~~arg ~\underset{\mathbf{X}}{min} \left \| \mathbf{Y}-\mathbf{\Phi} \mathbf{FX} \right \|_{F}^{2}+
\frac{\eta _{1}}{2}\left \| \mathbf{W^{\textit{j}}-\mathbf{\Psi}X-B_{1}^{\textit{j-1}}} \right \|_{F}^{2}+\\
\frac{\eta _{2}}{2}\left \| \mathbf{Z^{\textit{j}}-XD-B_{2}^{\textit{j-1}}} \right \|_{F}^{2},
\label{eq:no10DTSR}
\end{split}
\end{equation}
where \textit{j} is the iteration number. \textit{P}1 and \textit{P}2 subproblems minimize objective functions over \textbf{W} and \textbf{Z}, respectively, with fixed \textbf{X}. \textit{P}3 minimizes the objective function over \textbf{X} with fixed \textbf{W} and \textbf{Z}.  Above three subproblems are solved iteratively along with the update of Lagrange multipliers $\textbf{B}_{1}$ and $\textbf{B}_{2}$. The complete algorithm is summarized in Algorithm \ref{algorithm1DTSR}, while the solution of each subproblem is explained in the next few subsections. 

\begin{algorithm}
\caption{Pseudo code of the proposed DTSR method}
\label{algorithm1DTSR}
\begin{algorithmic}[1]
\item{Initialize $\lambda _{1}$, $\lambda _{2}$, $\eta _{1}$, $\eta _{2}$, $\textbf{B}_{1}^{0}$, $\textbf{B}_{2}^{0}$, $\textbf{X}^{0}$, \textit{j}=1 }
\While{convergence criteria not met}
\item{~~~~\textit{P}1-subproblem}
\begin{equation*}              
\mathbf{W}^{j}=arg ~\underset{\mathbf{W}}{min} ~~\lambda _{1}\left \| \mathbf{W} \right \|_{1 }+ \frac{\eta _{1}}{2}\left \| \mathbf{W-\mathbf{\Psi}X^{\textit{j}-1}-B_{1}^{\textit{j}-1}} \right \|_{F}^{2}.
\end{equation*}

\item{~~~~\textit{P}2-subproblem}
\begin{equation*} 
\mathbf{Z}^{j}=arg~ \underset{\mathbf{Z}}{min} ~~\lambda _{2}\left \|  \mathbf{Z} \right \|_{1}+\frac{\eta _{2}}{2}\left \| \mathbf{Z-X^{\textit{j}-1}D-B_{2}^{\textit{j}-1}} \right \|_{F}^{2}.
\end{equation*}

\item{~~~~\textit{P}3-subproblem}
\begin{equation*} 
\begin{split}
\mathbf{X}^{j}=arg ~\underset{\mathbf{X}}{min} \left \| \mathbf{Y}-\mathbf{\Phi} \mathbf{FX} \right \|_{F}^{2}+
\frac{\eta _{1}}{2}\left \| \mathbf{W^{\textit{j}}-\mathbf{\Psi}X-B_{1}^{\textit{j}-1}} \right \|_{F}^{2}\\+
\frac{\eta _{2}}{2}\left \| \mathbf{Z^{\textit{j}}-XD-B_{2}^{\textit{j}-1}} \right \|_{F}^{2}.
\end{split}
\end{equation*}

\item{~~~~Lagrange multipliers update}
\begin{equation*} 
\begin{split}
\textbf{B}_{1}^{j}=\textbf{B}_{1}^{j-1}+\mathbf{\Psi}\textbf{X}^{j}-\textbf{W}^{j}.\\
\textbf{B}_{2}^{j}=\textbf{B}_{2}^{j-1}+\textbf{X}^{j}\textbf{D}-\textbf{Z}^{j}.
\end{split}
\end{equation*}

\item{~~~~\textit{j=j}+1}

\EndWhile
\end{algorithmic}
\end{algorithm}

\subsubsection{\textbf{P1 and P2 Subproblems}}

The first two subproblems are $l^{1}$ minimization problems. For any $l^{1}$ minimization problem such as
\begin{equation}
\underset{\mathbf{P}}{min}~\alpha \left \| \mathbf{ P} \right \|_{1}+\frac{\beta }{2}\left \| \mathbf{P-Q} \right \|_{F}^{2},
\label{eq:no11DTSR}
\end{equation}
where $\textbf{P}, \textbf{Q}\in\mathbf{R}^{n\times T}$ and  $\alpha ,\beta > 0$, the solution is  \cite{Aggarwal2016}
\begin{equation}
\mathbf{P}=Soft(\mathbf{Q},2\frac{\alpha}{\beta }\mathbf{A}),
\label{eq:no12DTSR}
\end{equation}
where \textbf{A} is a matrix containing all ones and \textbf{Q} on the right hand side in the above equation is an initial estimate of \textbf{P}. The definition of `\textit{Soft}' is 

\begin{equation}
Soft(\mathbf{Q},\nu \mathbf{A})=sgn(\mathbf{Q})\otimes max\left \{ 0,\left | \mathbf{Q} \right |-\nu\mathbf{A} \right \},
 \label{eq:no13DTSR}
\end{equation}
where  $\nu=\frac{\alpha}{\beta }$, $\otimes$ denotes the element-wise product, and $|\textbf{Q}|$ denotes a matrix with absolute values of \textbf{Q}. \textbf{A} in the above equation ensures soft thresholding on all elements of \textbf{Q}. For the nonzero elements of \textbf{Q}, $sgn(\textbf{Q}) = \textbf{Q} ./\left | \textbf{Q}\right |$, otherwise sgn(\textbf{Q}) = 0. 

Hence, the closed form solution of  $\textbf{W}$ at the iteration number $\textit{j}$ in the \textit{P}1-subproblem is

\begin{equation}
\mathbf{W}^{j}=Soft((\mathbf{\Psi}\mathbf{X}^{j-1}+\mathbf{B}_{1}^{j-1}),2\frac{\lambda _{1}}{\eta _{1}}\mathbf{A}).
\label{eq:no14DTSR}
\end{equation}

Once $\textbf{W}$ at iteration $\textit{j}$ is estimated from the subproblem \textit{P}1, the next step is to estimate $\textbf{Z}$ from the subproblem \textit{P}2. Using $\textbf{X}^{j-1}$ and $\textbf{B}_{2}^{j-1}$ from the previous iteration, $\textbf{Z}^{j}$ can be obtained via a closed form as

\begin{equation}
\mathbf{Z}^{j}=Soft((\mathbf{X}^{j-1}D+\mathbf{B}_{2}^{j-1}),2\frac{\lambda _{2}}{\eta _{2}}\mathbf{A}).
\label{eq:no15DTSR}
\end{equation}

\subsubsection{\textbf{P3 Subproblem}}
This subproblem is quadratic. It can be efficiently solved using the conjugate gradient algorithm.  We used the line search conjugate gradient algorithm as used in \cite{lingala2011}.

\subsubsection{\textbf{Update of Lagrange Multiplier Variables}}
Last step is the update of Lagrange multipliers that is explained in Algorithm \ref{algorithm1DTSR}. Lagrange multipliers help in achieving convergence in the subsequent iterations. In this algorithm, convergence is checked either by comparing convergence of the objective function in \eqref{eq:no4DTSR} with a threshold or with the maximum number of iterations reached. 

\section{Results}

\subsection{Implementation Details}
\subsubsection{Temporal Sparsity Domain}
\begin{figure*}[!ht]
\begin{subfigure}[b]{0.5\textwidth}
\includegraphics[width=\textwidth]{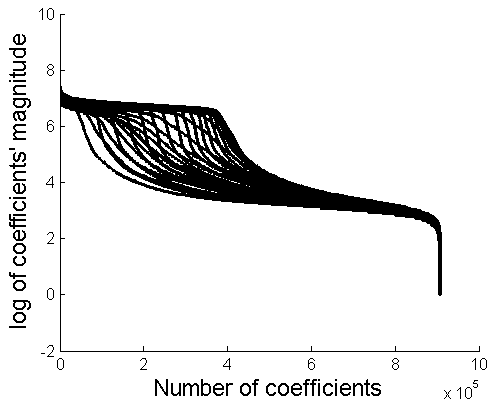}
 \caption{}
\end{subfigure}
\begin{subfigure}[b]{0.5\textwidth}
\includegraphics[width=\textwidth]{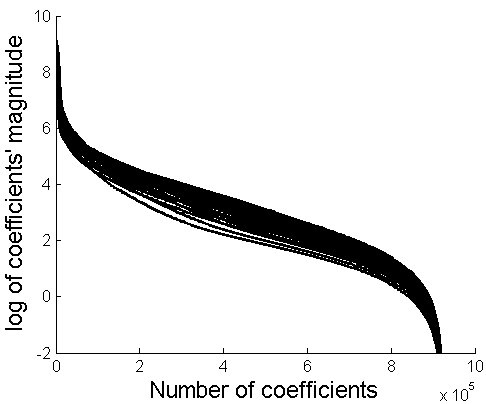}
 \caption{}
\end{subfigure}

\begin{subfigure}[b]{0.5\textwidth}
\includegraphics[width=\textwidth]{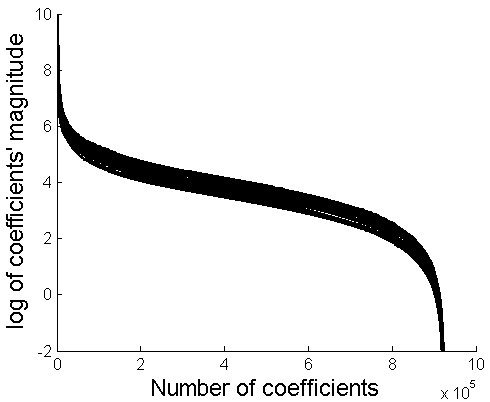}
 \caption{}
\end{subfigure}
\begin{subfigure}[b]{0.5\textwidth}
\includegraphics[width=\textwidth]{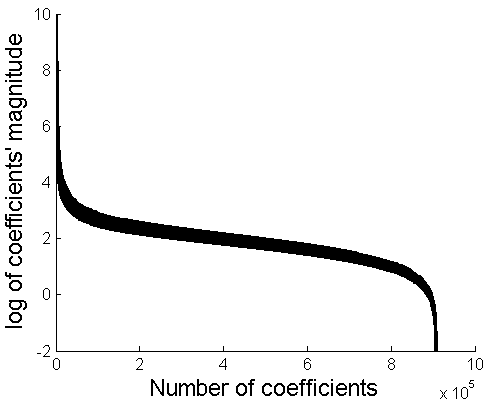}
 \caption{}
\end{subfigure}
\caption{Illustration of increased sparsity of fMRI data in the temporal Fourier Transform domain. One subject's sorted log magnitude values of coefficients of matrix \textbf{X} obtained using: (\textit{a}) no transform; (\textit{b}) 3-level dB4 wavelet in spatial direction; (\textit{c}) 2-D Discrete Cosine Transform in spatial directions; (\textit{d}) 1-D Discrete Fourier Transform along the rows of \textbf{X}.} \label{figure:no3DTSR}
\end{figure*}

In general, sparsity is imposed in the transform domain $\mathbf{\Psi}$ as shown in \eqref{eq:no2DTSR} and \eqref{eq:no4DTSR}. Recently, in our previous work \cite{Aggarwal2016}, we observed that fMRI data is more sparse in the temporal Fourier domain, i.e., in the Fourier domain of every voxel's time series. Hence, the matrix resulting by computing the Fourier transform of \textbf{X} along every row leads to a temporal Fourier transformed matrix that is sparse. In order to demonstrate this, we plot the sorted temporal Fourier transformed coefficients of matrix \textbf{X} corresponding to the middle slice of one subject (Refer to Fig.\ref{figure:no3DTSR}).

From these figures, we observe that fMRI data is sparser in the temporal frequency domain. Hence, in \eqref{eq:no4DTSR}, we consider $\mathbf{\Psi}$ as the corresponding temporal Fourier domain sparsifying matrix, i.e., it computes the Fourier transform along every row of \textbf{X}. 

\subsubsection{Retrospective Undersampling}
In this paper, undersampled $k-t$ space data \textbf{Y} is acquired by retrospective undersampling of the Fourier transform of the fully available data. Radial sampling patterns are used to undersample available $k$-space data as used in \cite{radial,Aggarwal2017,Aggarwal2016}.  These patterns represent different sampling masks used for one brain slice of size $n_{x}\times n_{y}$ at one time point with zeros at non-sampled locations and ones at sampled locations. Sensing matrix $\mathbf{\Phi}$ in \eqref{eq:no4DTSR} is constructed by stacking sampling masks of all \textit{T} time points.

\subsubsection{Regularization Parameters}

The proposed DTSR method requires seven parameters $\lambda_{1}$, $\lambda_{2}$, $\eta_{1}$, $\eta_{2}$, $\textbf{B}_{1}^{0}$, $\textbf{B}_{2}^{0}$, and $\textbf{X}^{0}$ to be initialized as shown in Algorithm \ref{algorithm1DTSR}.  $\eta_{1}$ and $\eta_{2}$ are initialized as $10^{-2}$. Lagrange multipliers $\textbf{B}_{1}^{0}$ and  $\textbf{B}_{2}^{0}$ are initialized to matrices containing all one's. We used $L$-curve method to initialize $\lambda_{1}$ and $\lambda_{2}$ \cite{Lcurve} based on the maximization of peak signal-to-noise ratio (PSNR) compared to the ground truth (fully available dataset). With this method, we arrive at the following values: $\lambda_{1}$=$\lambda_{2}$=0.5. The fMRI data matrix $\textbf{X}^{0}$ in subproblem \textit{P}1 is initialized using the crude initial estimate obtained via direct inverse Fourier transform (IFT). Direct IFT method computes IFT of given $k-t$ space data $\textbf{Y}$ as shown below:
\begin{equation}
\textbf{X}^{0}=IFT(\mathbf{Y}).
\label{eq:no16DTSR}
\end{equation}
\subsection{Related Reconstruction Methods}
We compare results of the proposed DTSR method with other offline fMRI reconstruction methods including CSWD \cite{Holland2013}, HSPARSE \cite{Fang2015}, \textit{k-t} FASTER \cite{Chiew2016}, and LR+S \cite{Singh2015}.
Below we present brief overview of each of the reconstruction methods implemented. We also provide regularization parameter values used in the simulation of these methods.

\subsubsection{CS with Wavelet Domain Sparsity (CSWD) \cite{Holland2013}} In this method, compressive sensing based reconstruction of fMRI data is carried out assuming the fMRI data to be spatially sparse in the wavelet domain \cite{Holland2013}. Hence, fMRI reconstruction is done by using the optimization framework as explained in \eqref{eq:no2DTSR}. Here, $\mathbf{\Psi}$ is a wavelet matrix operator. We used Daubechies orthogonal wavelet `db4' (filter lengths 8) with 3-level decomposition as used in \cite{Holland2013}. We used non-linear conjugate gradient method to solve CSWD \cite{lustig2007sparse}. We used the default value of $\lambda_{1}$=0.1 as specified in this method. 

 \subsubsection{HSPARSE Method \cite{Fang2015}} 
The HSPARSE method reconstructs fMRI data assuming data matrix \textbf{X} to be sparse in both the temporal and spatial domains. This method is implemented by solving the below optimization problem \cite{Fang2015}:

\begin{equation}
\hat{\mathbf{X}}=arg ~\underset{\mathbf{X}}{min} \left \| \mathbf{Y}-\mathbf{\Phi} \mathbf{FX} \right \|_{F}^{2}+\lambda_{3}\left \| \mathbf{\Psi_{t}}\mathbf{X} \right \|_{1 }+\lambda_{4}\left \| \mathbf{\Psi_{s}}\mathbf{X} \right \|_{1 },
\label{eq:no17DTSR}
\end{equation}
where $\lambda_{3}$ and $\lambda_{4}$ are regularization parameters and, $\mathbf{\Psi_{t}}$ and $\mathbf{\Psi_{s}}$ are the temporal and spatial domain sparsifying basis, respectively. We chose discrete cosine transform (DCT) for both the temporal and the spatial sparsity as used in \cite{Fang2015} and $\lambda_{3}$=0.5 and $\lambda_{4}$=0.1 in \eqref{eq:no17DTSR} for simulation. We used the non-linear conjugate gradient method to solve HSPARSE \cite{lustig2007sparse}. 
 
\subsubsection{k-t FASTER Method \cite{Chiew2016}} 
\textit{k-t} FASTER method reconstructs fMRI data assuming data matrix \textbf{X} to be low rank. This method is implemented by solving the below optimization problem \cite{Chiew2016}:
\begin{equation}
\hat{\mathbf{X}}=arg ~ \underset{\mathbf{X}}{min} \left \| \mathbf{Y}-\mathbf{\Phi} \mathbf{FX} \right \|_{F}^{2}~~s.t ~rank(\mathbf{X})=r,
\label{eq:no18DTSR}
\end{equation}
where \textit{r} is the pre-defined rank of \textbf{X}. In \textit{k-t} FASTER \cite{Chiew2016}, hard thresholding is applied on the singular values of data matrix \textbf{X} as explained below. First, singular value decomposition (SVD) of an initial crude estimate of matrix $\textbf{X}^{0}$ is computed 
\begin{equation}
\mathbf{X}^{0}=\mathbf{USV^{\mathit{T}}}.
\label{eq:no19DTSR}
\end{equation}
Next, hard thresholding is applied on the singular values contained in $\textbf{S}$ as 

\begin{equation}
 \hat{s_i}=\left\{
\begin{matrix}
|s_i|-\mu & i \leq r  \\  
0  & i > r 
\end{matrix} \right.
\label{eq:no20DTSR}
\end{equation}
where $\mu$ is a constant, $s_{i}$ is the $i^{th}$ singular value of \textbf{S}, and $\hat{s_i}$ is updated singular value after hard thresholding. 

The value of constant $\mu$ is chosen to be 0.5 as used in \cite{Chiew2016}. In simulation, rank $r$ is considered to be equal to the number of time frames. This value provided least normalized mean square error (NMSE) between the reconstructed and the original fMRI data considered in this work.

\subsubsection{Low Rank plus Sparse (LR+S) Method \cite{Singh2015}} This method reconstructs fMRI data using low rank and sparse matrix decomposition and hence, is solved using the following optimization framework:
\begin{equation}
\hat{\mathbf{L}},\hat{\mathbf{S}}=arg~ \underset{\mathbf{L,S}}{min} \left \| \mathbf{Y}-\mathbf{\Phi} \mathbf{F(L+S)} \right \|_{F}^{2}+\lambda _{5}\left \| \mathbf{L} \right \|_{\ast }+\lambda _{6}\left \|  \mathbf{S} \right \|_{1},
\label{eq:no21DTSR}
\end{equation}
where $ \lambda _{5}$ and $ \lambda _{6}$ are the regularization parameters. The fMRI data matrix $\textbf{X}$ is reconstructed as: 
\begin{equation}
\hat{\mathbf{X}}=\hat{\mathbf{L}}+\hat{\mathbf{S}}.
\label{eq:no22DTSR}
\end{equation}	
We empirically selected $\lambda _{5}=300$ and $\lambda _{6}=0.5$ that provided us the minimum NMSE.

The default number of iterations used in the non-linear conjugate algorithm in CSWD \cite{Holland2013} and HSPARSE \cite{Fang2015} is 20 \cite{lustig2007sparse}. Hence, for other reconstruction methods including DTSR, we have set the maximum number of iterations (required in optimization) to 20 and the convergence threshold of objective function to $10^{-5}$. Since LR+S \cite{Singh2015} method converges in more number of iterations, we predefined the maximum number of iterations to 100 for this method.

\subsection{Quantitative Analysis}
In this section, we compare results of the proposed DTSR against the existing CS-based fMRI reconstruction methods. The Normalized Mean Square Error (NMSE), PSNR, and Structural Similarity Index (SSIM) are used as reconstruction quality assessment metrics in this paper.

NMSE and PSNR are two well known reconstruction quality assessment metrics. Given a reference brain slice $\textbf{x}_{t}$  at time point $t$ and it's reconstructed estimate $\hat{\textbf{x}}_{t}$ ($t^{th}$ column of $\hat{\textbf{X}}$), NMSE is calculated as

\begin{equation}
\label{eq:no23DTSR}
\textup{NMSE}=\left \| \mathbf{x}_{t}-\hat{\mathbf{x}_{t}} \right \|_{2}/\left \| \mathbf{x}_{t} \right \|_{2},
\end{equation}
where $\left \|  .\right \|_{2}$ denotes $l^{2}$ norm. Similarly PSNR is calculated as

\begin{equation}
\label{eq:no24DTSR}
\textup{PSNR}=20\textup{log}_{10}\frac{255}{\frac{1}{n_{x}n_{y}} \left \| \mathbf{x}_{t}-\hat{\mathbf{x}_{t}} \right \|_{2}},
\end{equation}
where $n_{x} \times n_{y}$ denotes the size of brain slices. In this work, one slice is being reconstructed simultaneously over all time points. Hence, NMSE and PSNR are calculated using \eqref{eq:no23DTSR} and \eqref{eq:no24DTSR} for \textit{T} number of time points for a given slice and are subsequently time-averaged. In the following text, NMSE and PSNR  signify average NMSE and average PSNR. In addition, SSIM \cite{Wang2009} is also used to measure the reconstruction quality. SSIM is known to be a better metric than NMSE and PSNR \cite{Wang2009}.

Reconstructed (average) NMSE, (average) PSNR, and (average) SSIM results over all 33 slices of one subject are presented in Table \ref{Table1DTSR} and \ref{Table2DTSR}. Results are tabulated at 6, 12, and 24 number of radial sampling lines. From these results, we observe that the proposed DTSR performs consistently better than the existing reconstruction methods. This is to note that the proposed DTSR method assumes double sparsity in the temporal domain. Hence, it yields better results compared to the existing methods and also reconstructs fMRI data quite efficiently at lower number of radial sampling lines.

\begin{table*}[!htbp]
\caption{Reconstruction results with different methods on real fMRI dataset$^{a}$} \label{Table1DTSR}
\begin{tabular}{|c|c|c|c|c|c|c|}
\hline 
 &  & \textbf{NMSE }&  &  & \textbf{PSNR} &   \tabularnewline
\hline 
\hline 
\textbf{Method }  & \textbf{6 lines} & \textbf{12 lines} & \textbf{24 lines} & \textbf{6 lines }& \textbf{12 lines} & \textbf{24 lines } \tabularnewline
\hline 
CSWD \cite{Holland2013} & 0.2574 & 0.1725 & 0.1153 & 4.898 & 8.3784 & 11.8714 \tabularnewline
\hline 
HSPARSE \cite{Fang2015} & 0.1801 & 0.0863 & 0.0573 & 8.013 & 14.401 & 17.955 \tabularnewline
\hline 
k-t FASTER \cite{Chiew2016} & 0.2566 & 0.1725 & 0.1155 & 4.929 & 8.379 & 11.8574 \tabularnewline
\hline 
LR+S \cite{Singh2015} & 0.1767 & 0.0963 & 0.0544 & 8.175 & 13.438 & 19.071 \tabularnewline
\hline 
Proposed DTSR & \textbf{0.0471} & \textbf{0.0382 }& \textbf{0.036} & \textbf{19.64 }&\textbf{ 21.48} & \textbf{21.99} \tabularnewline
\hline 
Proposed DTSR with $\lambda_{2}=0$ & 0.1264 & 0.0875& 0.0751 & 11.45&19.63 &20.93 \tabularnewline
\hline 
Proposed DTSR with $\lambda_{3}=0$ & 0.0974 & 0.0765 & 0.0479 & 18.22 & 20.12 & 21.15 \tabularnewline
\hline
\end{tabular}
\begin{flushleft} $^{a}$ Dataset- Beijing\_Zang resting state fMRI data, subject no. 1, results averaged over all slices and averaged over all time points.
\end{flushleft}
\vspace{40mm}
\end{table*}

\begin{table*}
\caption{SSIM performance on real fMRI data$^{a}$ reconstructed with different methods} \label{Table2DTSR}
\centering
\begin{tabular}{|c|c|c|c|}
\hline 
 &  & \textbf{SSIM} & \tabularnewline
\hline 
\hline 
\textbf{Method }  & \textbf{6 lines} & \textbf{12 lines} & \textbf{24 lines} \tabularnewline
\hline 
CSWD \cite{Holland2013} & 0.3455 & 0.5248 & 0.7058\tabularnewline
\hline 
HSPARSE \cite{Fang2015} & 0.5241 & 0.8045 & 0.879\tabularnewline
\hline 
k-t FASTER \cite{Chiew2016} & 0.3475 & 0.5237 & 0.7052\tabularnewline
\hline 
LR+S \cite{Singh2015} & 0.5863 & 0.7878 & 0.9142\tabularnewline
\hline 
Proposed DTSR & \textbf{0.9149} & \textbf{0.9323 }& \textbf{0.9441}\tabularnewline
\hline 
\end{tabular}
\begin{flushleft} $^{a}$ Dataset- Beijing\_Zang resting state fMRI data, subject no. 1, results averaged over all slices and averaged over all time points.
\end{flushleft}
\end{table*}

We observe that NMSE and PSNR are consistently very high at all acceleration factors considered including lower acceleration factors. This implies that we can reconstruct fMRI data by sampling much lesser measurements in $k-t$ space with the proposed DTSR method compared to the other methods. Hence, higher acceleration is possible with the DTSR method that in turn will decrease the fMRI acquisition time.

\begin{figure*}
 \centering
    \includegraphics[scale=0.8]{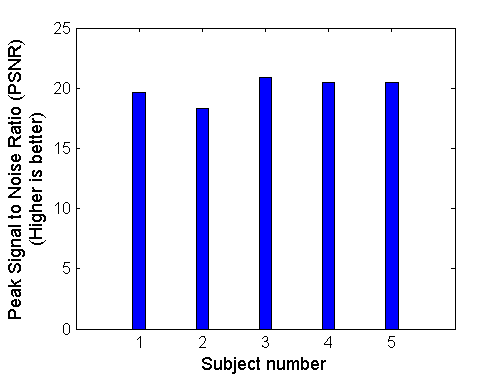}
\caption{Illustration of reconstruction performance (in terms of average PSNR) over different subjects}  \label{figure:no4DTSR}
\end{figure*}

Fig.\ref{figure:no4DTSR} presents (average) PSNR results over all slices of fMRI data of 5 subjects. Six radial lines are used for undersampling the \textit{k}-space data (12.856 acceleration factor). These results indicate that the proposed DTSR method is robust to subject variability and that the results are reproducible across subjects.

\begin{figure*}
\centering
\includegraphics[scale=0.48, trim = 20 0 0 0]{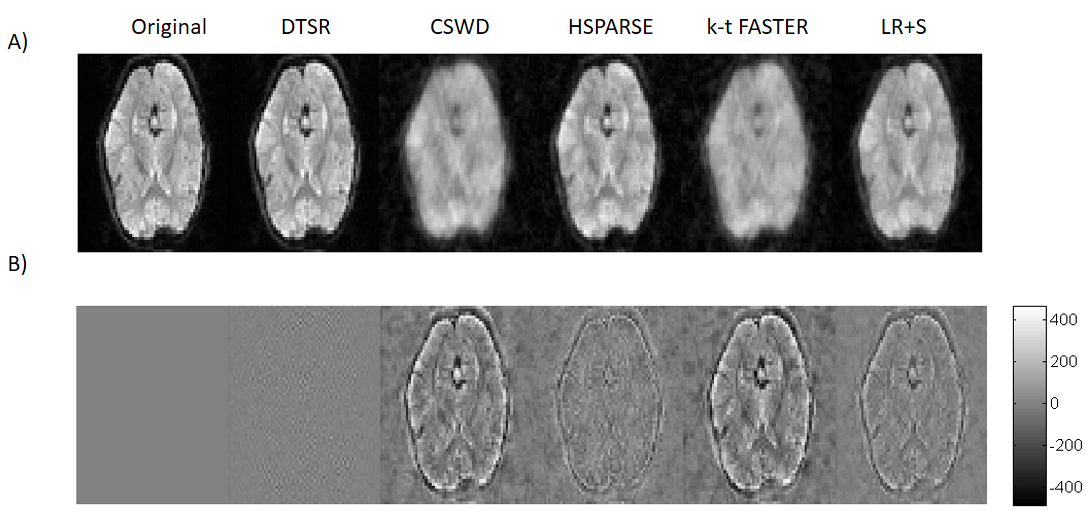}
 \caption{Reconstructed middle slice (slice no.16 (total slices = 33)) using different reconstruction methods with 6 radial lines. Left to right: original; proposed DTSR; CSWD; HSPARSE, k-t FASTER; LR+S. A) The reconstructed middle slice obtained using different methods on the fMRI data of one randomly selected subject and at one random time point. B) Difference images (Ground truth $–$ Reconstructed) corresponding to the original slice presented in column one of subfigure-(A)}
 \label{figure:no5DTSR}
\end{figure*}

Furthermore, to visually compare the data reconstructed  using different methods, we present reconstructed middle slice (slice no.16 (total slices = 33)) in Fig.\ref{figure:no5DTSR} on one random subject and at one random time point using six radial lines. From  Fig.\ref{figure:no5DTSR}, we observe that the reconstruction quality
with the proposed DTSR method is superior compared to the existing methods.

\subsection{Qualitative Analysis}
In this section, we compare and evaluate the reproducibility of brain RSNs constructed using the proposed DTSR based reconstruction fMRI data and using the fully available whole brain resting fMRI dataset. Spatial Independent Component Analysis (ICA) of the reconstructed fMRI data and the original fully available fMRI dataset is performed via GIFT toolbox \footnote{\url{https://www.nitrc.org/projects/gift}}. ICA is a data driven method that has been widely used in resting state fMRI to recover the set of spatially independent brain RSNs \cite{calhoun2013,Allen2012,kiviniemi2009,abou2010,smith2009}.

\subsubsection{Data Preprocessing}
fMRI data suffers from low SNR and hence, needs to be preprocessed before analysis. Preprocessing is performed using SPM12 (Statistical Parametric Mapping)\footnote{\url{http://www.fil.ion.ucl.ac.uk/spm/software/spm12/}} and Matlab. The fMRI brain volumes are slice-time corrected using the middle slice (16, total slices=33) as a reference, realigned to the mean image, spatially normalized onto the Montreal Neurological Institute (MNI) space (3-mm isotropic voxels), and are spatially smoothed with a Gaussian kernel (Full Width Half Maximum (FWHM)=4 mm). 

\subsubsection{The ICA Model}
For the sake of completeness, ICA model as used in fMRI is briefly discussed in this subsection. Consider matrix $\textbf{S}\in \mathbb{R}^{T\times V}$, where  \textit{T} is the number of time points and \textit{V} is the total number of voxels. After ICA, \textbf{S} can be expressed as:
\begin{equation}
\textbf{S}=\textbf{MN},
\label{eq:no25DTSR}
\end{equation}
where \textbf{M} is the $T\times C$ mixing matrix and \textbf{N} is the $C\times V$ source matrix. \textit{C} is the total number of spatially independent component. Each row of source matrix \textbf{N} represents one spatially independent component and the corresponding column of the mixing matrix \textbf{M} represents time course of that independent component. The goal of spatial ICA is to model fMRI data as a mixture of maximally  independent spatial components. 

In this paper, the InfomaxICA algorithm \cite{bell1995} is used to obtain ICA components and the corresponding time courses. It is one of the most popular ICA algorithms that is used in fMRI data analysis \cite{esposito2002}. The number of spatially independent components \textit{C} is predefined to 100 in accordance with the previous studies \cite{Allen2012,kiviniemi2009,abou2010,smith2009}. High number of components facilitate good segregation of cortical and subcortical brain functional networks \cite{Allen2012}. 

To identify brain RSNs among 100 spatially independent components, spatial distribution of each component can be identified by spatial overlapping with the available template images of brain RSNs. We identified 51 ICs from the mean maps of all 20 fully available fMRI subjects after removing the artifact components. These ICs can be broadly categorized into 10 RSNs: 1. Visual Network (VN), 2. SomatoMotor Network (SMN), 3. Limbic Network (LN), 4. Dorsal Attention Network (DAN), 5. Ventral Attention Network (VAN), 6. Default Mode Network (DMN), 7. Frontoparietal Network (FPN), 8. Temporal + Frontal Network (TFN), 9. Subcortical Network (SCN), and 10. Cerebellar Network (CN).

\begin{figure*}
\includegraphics[width=\textwidth]{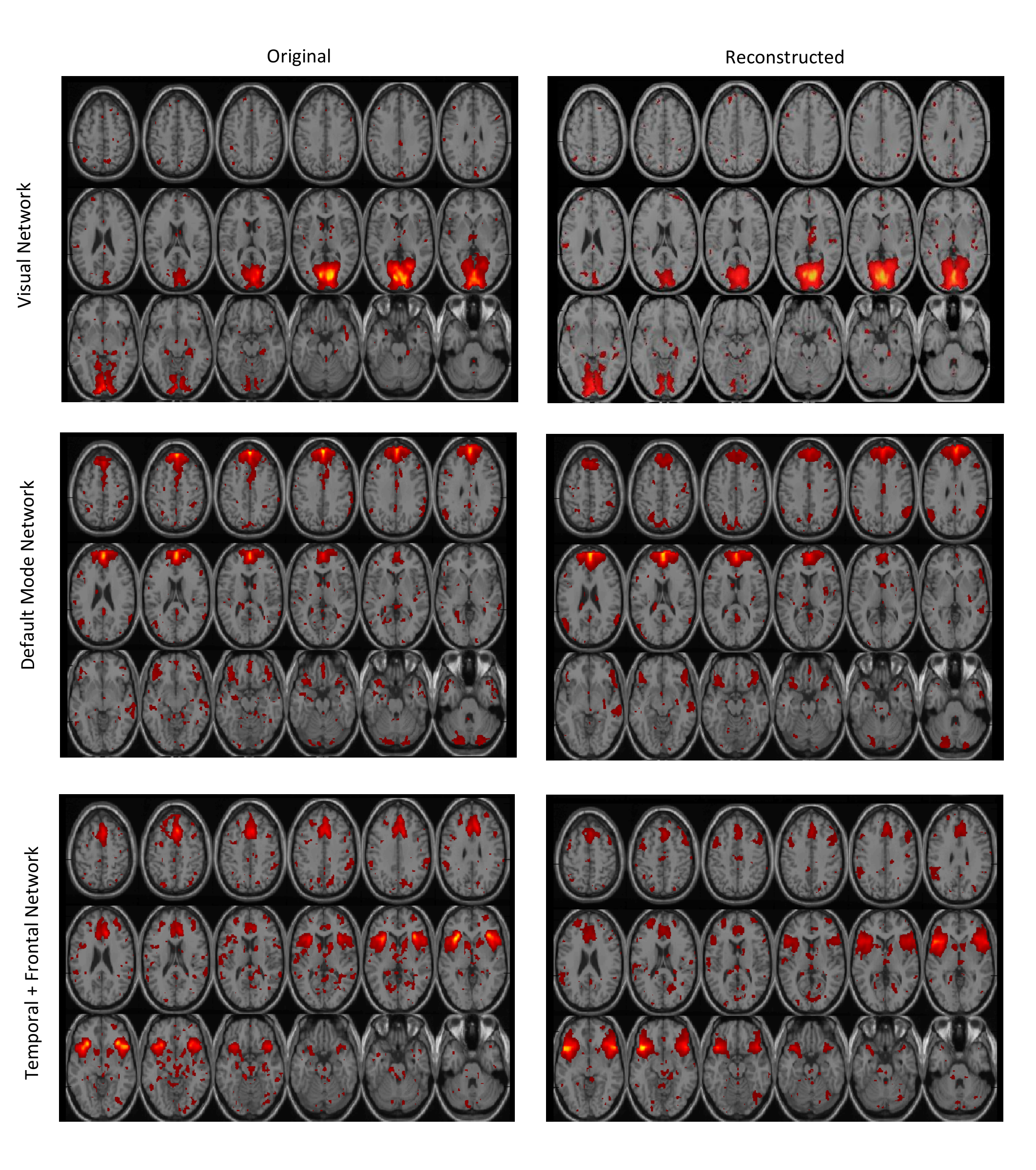}
 \caption{Axial view of spatial maps of various RSNs where the left part of each figure represents networks identified using the fully available (original) data and the right part represents networks identified using the DTSR reconstructed data. Each row corresponds to results on one RSN. Number in brackets below each image represents independent component (IC) number obtained after group ICA.}\label{figure:no6DTSR}
\end{figure*}

\begin{figure*}
\includegraphics[width=\textwidth]{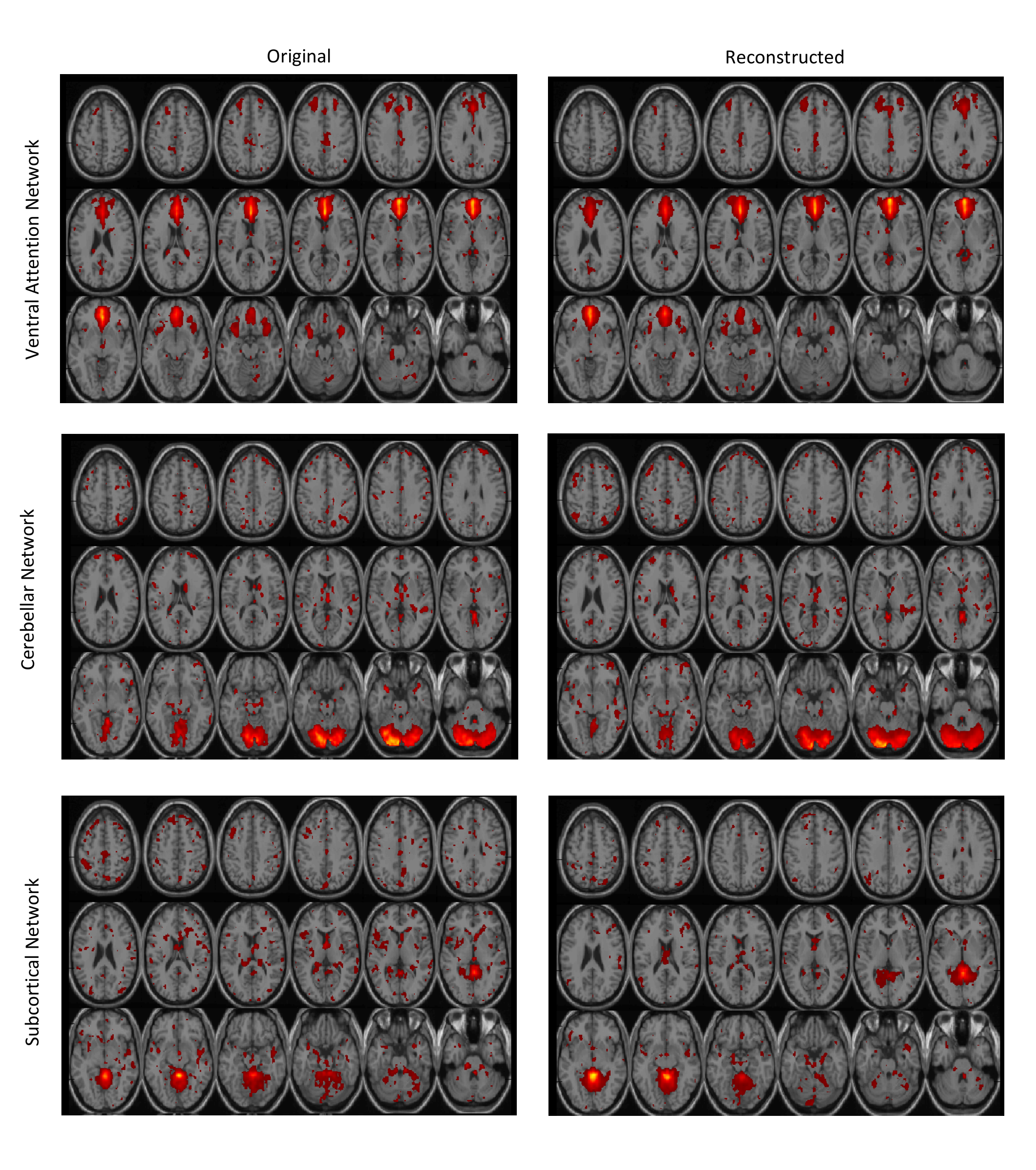}
 \caption{Axial view of spatial maps of various RSNs where the left part of each figure represents networks identified using the fully available (original) data and the right part represents networks identified using the DTSR reconstructed data. Number in brackets below each image represents independent component (IC) number obtained after group ICA.}\label{figure:no7DTSR}
\end{figure*}

On the fMRI data reconstructed using the proposed DTSR method, we identified 56 ICs from the mean maps of reconstructed fMRI data of all 20 subjects. We manually arranged these ICs into various RSNs stated above. The spatial maps of some RSNs discovered by the fully available data and the DTSR reconstructed data are shown in Fig.\ref{figure:no6DTSR} to Fig.\ref{figure:no7DTSR}. Left part of each figure represents networks identified using the fully available data and the right part represents networks identified using the DTSR reconstructed data. It is clear that spatial activation maps of RSNs obtained from the reconstructed data overlap significantly with the RSNs of the fully available fMRI data. From now onwards, ICs of DTSR reconstructed data will be mentioned as simply DTSR ICs (DTSR-ICs) and ICs of fully sampled original data will be mentioned as raw ICs (RICs). It is noticed that DTSR-IC maps are more enhanced compared to the RIC maps. This is perhaps due to denoising inherent within the CS reconstruction framework.

\begin{figure*}
\includegraphics[width=\textwidth]{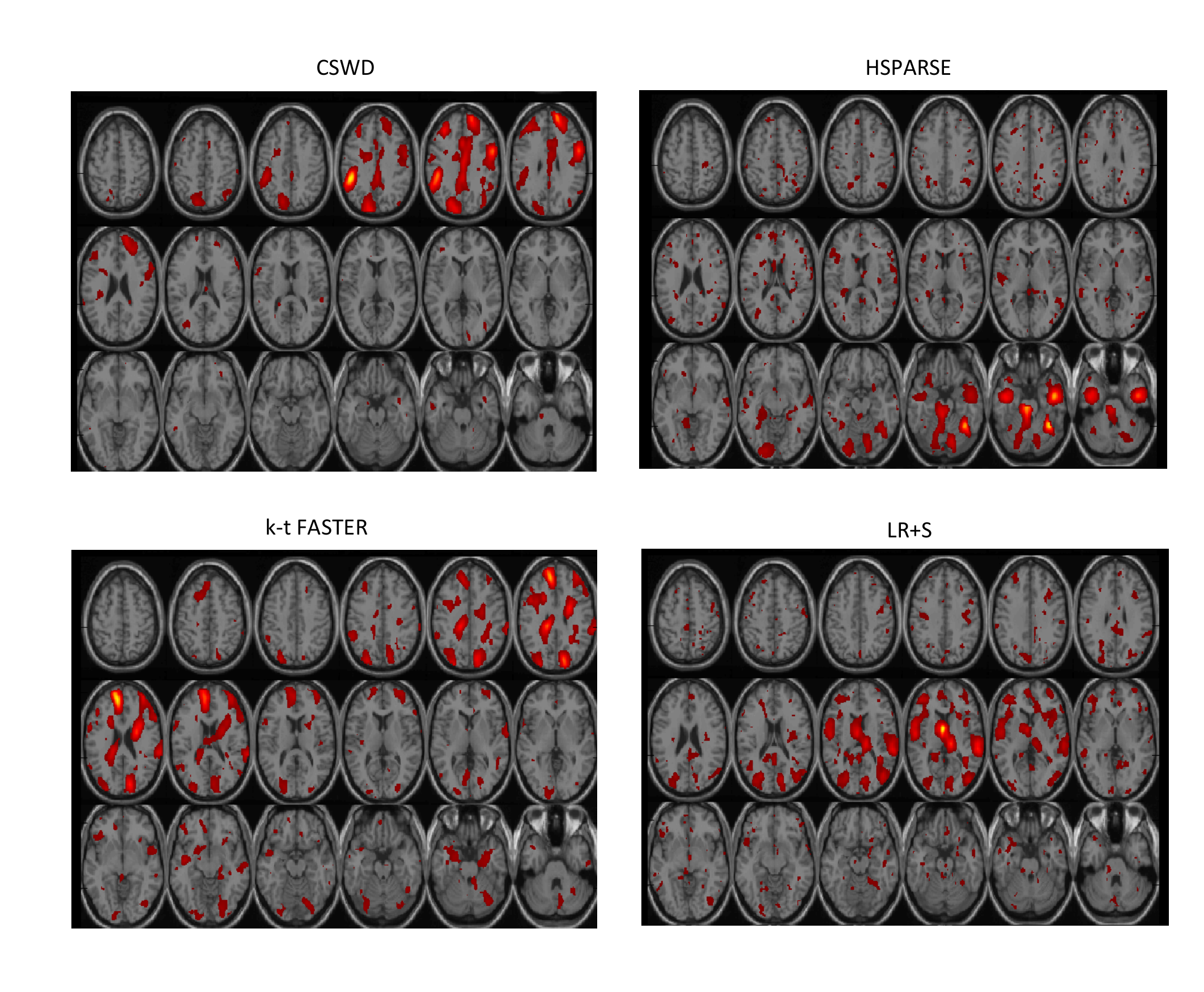}
 \caption{Axial view of one random spatial map obtained from the data reconstructed using existing methods.}\label{figure:no8DTSR}
\end{figure*}

From Fig.\ref{figure:no5DTSR}, we note that existing methods result in greater artifact compared to both the original data and the data obtained using the DTSR method. Further, we evaluated their performance in terms of reproducibility of ICA activation maps. One random activation map obtained using existing methods is presented in Fig.\ref{figure:no8DTSR}. From this figure, we observe that false maps are being detected with CSWD, HSPARSE, \textit{k-t} FASTER, and LR+S methods  that  can lead to misleading findings based on RSNs.

\begin{figure*}
\includegraphics[width=\textwidth]{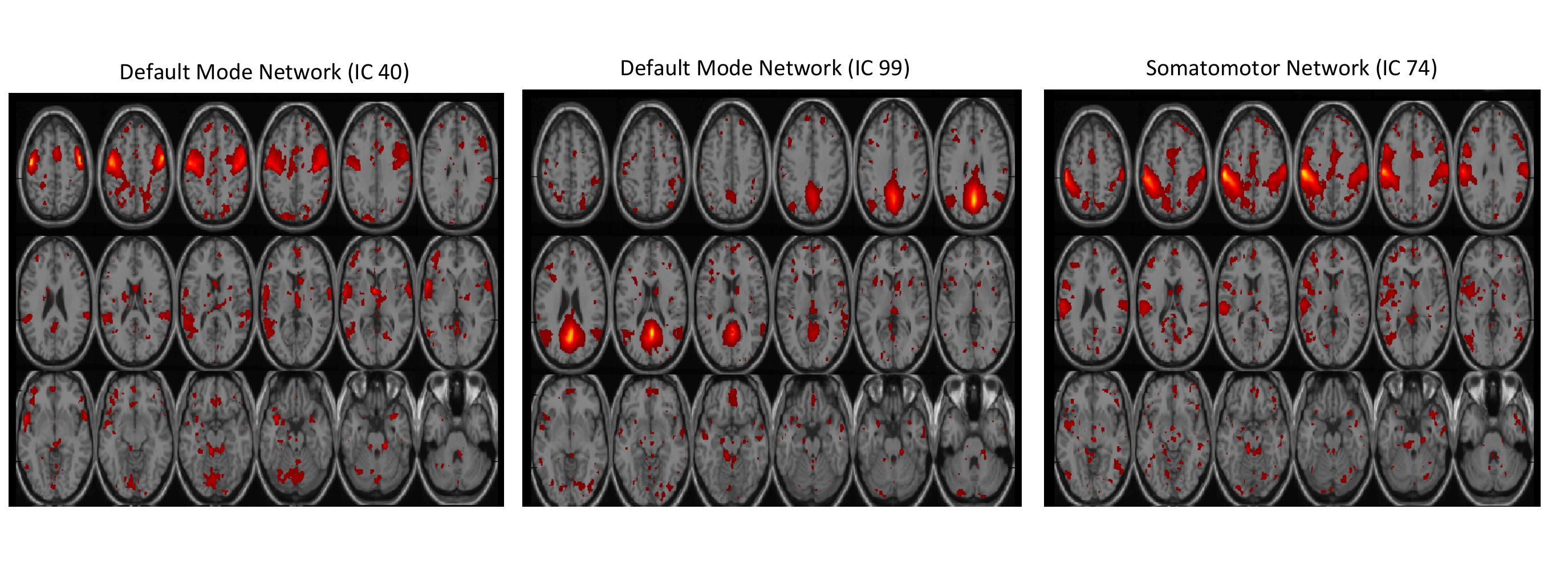}
 \caption{Axial view of spatial maps of various ICs obtained from the DTSR reconstructed data}\label{figure:no9DTSR}
\end{figure*}

This is to note that the DTSR-ICs of the reconstructed dataset matched with the RICs of fully sampled original data. Further, three additional ICs are observed with the DTSR reconstructed data that are not visible with the IC analysis of the original data. Fig.\ref{figure:no9DTSR} shows that these three ICs are actually the part of DMN and SMN. This shows that the data reconstructed using the proposed DTSR method has better RSN construction ability compared to the raw (original) data. This is to note that every RSN is observed to be present in all 20 subjects. This observation is in order because DTSR method denoises the data during reconstruction and hence, provides better network construction.

\section{Discussion}
This study proposes DTSR method that provides accelerated fMRI data reconstruction by imposing double temporal sparsity. Proposed method makes use of the advantage of repeated scanning of the same brain volume in fMRI to study brain's function. Scanning of the same brain volume over time brings similarity in the temporal direction that can be utilized via total variation based regularization. In addition to this, we also imposed conventional temporal sparsity in the proposed reconstruction framework and hence, name the proposed method as Double Temporal Sparsity based Reconstruction (DTSR).

Compressed sensing is used to recover undersampled data captured at different acceleration factors, although inherently weak BOLD signals in resting state fMRI prohibits the use of higher acceleration factors  \cite{Zong2014,Chiew2016}. However, in this work, we have been able to achieve robust and reproducible results with a higher acceleration factor of 12.856 (corresponding to 6 radial lines). In addition, results at lower acceleration factors are improved compared to the existing methods \cite{Holland2013,Fang2015,Chiew2016,Singh2015}. 

We have thoroughly evaluated the robustness and the feasibility of the proposed DTSR method both at the subject and the group levels of real fMRI data. The performance of DTSR method has been evaluated using retrospective undersampling of the fully available fMRI dataset. We compared the performance of the proposed method with other offline methods such as Compressed Sensing with Wavelet domain Sparsity (CSWD) \cite{Holland2013}, HSPARSE \cite{Fang2015}, \textit{k-t} FASTER \cite{Chiew2016}, and LR+S \cite{Singh2015}. Reconstruction reliability increases with increasing number of radial lines (lower acceleration factors). Table \ref{Table1DTSR} and \ref{Table2DTSR} show that error increases with decreasing number of radial lines. However, the proposed DTSR method is equally reliable at lower number of radial lines, i.e., at higher acceleration factors. Comparison with other methods shows that DTSR method produces lowest error (0.0471) at a high acceleration factor of 12.856 (corresponding to 6 radial lines). Further, results presented in Fig.5 demonstrate that the existing reconstruction methods produce greater artifact at this high acceleration factor.

The robustness and the reliability of the proposed method are further assessed using both quantitative and qualitative analyses on real fMRI data. Results from the quantitative analysis show that the NMSE of the proposed reconstruction method is less compared to that of the other reconstruction methods (Table \ref{Table1DTSR}). Thus, it is noted that the incorporation of total variation along with sparsity as used in the proposed DTSR method improves reconstruction results significantly. The proposed DTSR reconstruction method produces significantly high values of PSNR and SSIM, and lower values of NMSE compared to the other existing methods (Table \ref{Table1DTSR} and \ref{Table2DTSR}).

Two observations are in order from the qualitative results (Fig.\ref{figure:no5DTSR}-\ref{figure:no6DTSR}): (i) intrinsic resting state networks are consistent and comparable to the fully sampled fMRI data. Hence, crude estimate of regularization parameters is sufficient for reconstruction; and (ii) the fidelity of the proposed method in maintaining temporal information is established via consistency of results and via observation of all RSNs.

In summary, the proposed DTSR reconstruction method maintains temporal resolution even at higher acceleration factors without affecting the spatial resolution and can be used to provide accelerated fMRI reconstruction with accurate detection of intrinsic brain's Resting State Networks (RSNs). Moreover, the proposed DTSR method is able to improve BOLD sensitivity both at the individual and at the group level compared to the existing methods.

\subsection*{Limitations and Future Work}
This work has proposed DTSR reconstruction method that exhibits better accelerated reconstruction performance compared to the existing methods using radial sampling pattern. However, 3D radial Cartesian sampling grid is more practical from the point of view of compressed data acquisition in scanner \cite{Chiew2016}. In future, we will use realistic sampling patterns to undersample data.

Secondly, use of both parallel imaging and CS fMRI may give excellent high spatio-temporal fMRI quality such as that observed in the context of dynamic MRI \cite{tsao2003,tsao2005,huang2005}. Thus, further studies may be undertaken to enhance the performance of DTSR method by using it in conjunction with existing parallel fMRI imaging methods \cite{Hugger2011, Chaari2014}. 

Thirdly, CS has been widely used to improve fMRI acquisition time. Hence, using this in conjunction with other high spatial resolution techniques, such as super-resolution technique \cite{Reeth2012}  or simultaneous multi-slice imaging \cite{Feinberg2013},  may provide high spatio-temporal fMRI  \cite{Fang2015}. In the future, we will explore accelerated reconstruction from the perspective of improving spatial resolution. 

Fourth, time complexity of DTSR is high because it involves multiplication of large \textbf{X} and \textbf{D} matrices. In the future, we will use GPU based computation to lower the computational time complexity. 

Finally, further evaluation of the proposed method on prospective undersampled fMRI data will help to check its robustness in real scenarios. 

\section{Conclusions}
In this paper, we have introduced a novel accelerated fMRI reconstruction method that exploits the advantage of scanning the same brain volumes in fMRI over a number of time points via double temporal sparsity constraints. The proposed DTSR reconstruction method can be used to acquire high temporal resolution fMRI data in smaller times comparable to those of lower resolution fMRI data along with accurate detection of intrinsic resting state brain networks. The performance of the proposed method has been evaluated using retrospective undersampling of the fully available real fMRI dataset. Code of proposed DTSR method can be obtained from (\url{http://in.mathworks.com/matlabcentral/fileexchange/63768-dtsr-fmri-reconstruction}).

\section*{Acknowledgements}
The first author would like to thank Visvesvaraya research fellowship, Department of Electronics and Information Tech., Ministry of Comm. and IT, Govt. of India, for providing financial support for this work.

\section*{References}

\bibliography{referencefusedLassoV1}

\begin{thebibliography}{49}
\providecommand{\natexlab}[1]{#1}
\providecommand{\url}[1]{\texttt{#1}}
\providecommand{\urlprefix}{URL }
\expandafter\ifx\csname urlstyle\endcsname\relax
  \providecommand{\doi}[1]{doi:\discretionary{}{}{}#1}\else
  \providecommand{\doi}[1]{doi:\discretionary{}{}{}\begingroup
  \urlstyle{rm}\url{#1}\endgroup}\fi
\providecommand{\bibinfo}[2]{#2}

\bibitem[{Logothetis(2008)}]{fMRIref1}
\bibinfo{author}{N.~K. Logothetis}, \bibinfo{title}{What we can do and what we
  cannot do with f{MRI}}, \bibinfo{journal}{Nature}
  \bibinfo{volume}{453}~(\bibinfo{number}{7197}) (\bibinfo{year}{2008})
  \bibinfo{pages}{869--878}, ISSN \bibinfo{issn}{0028-0836}.

\bibitem[{Feinberg and Yacoub(2012)}]{Feinberg2012}
\bibinfo{author}{D.~A. Feinberg}, \bibinfo{author}{E.~Yacoub},
  \bibinfo{title}{The rapid development of high speed, resolution and precision
  in f{MRI}}, \bibinfo{journal}{NeuroImage}
  \bibinfo{volume}{62}~(\bibinfo{number}{2}) (\bibinfo{year}{2012})
  \bibinfo{pages}{720--725}, ISSN \bibinfo{issn}{1053-8119}.

\bibitem[{Ogawa et~al.(1990)Ogawa, Lee, Kay, and Tank}]{Ogawa1990}
\bibinfo{author}{S.~Ogawa}, \bibinfo{author}{T.~M. Lee}, \bibinfo{author}{A.~R.
  Kay}, \bibinfo{author}{D.~W. Tank}, \bibinfo{title}{Brain magnetic resonance
  imaging with contrast dependent on blood oxygenation.},
  \bibinfo{journal}{Proc Natl Acad Sci USA}
  \bibinfo{volume}{87}~(\bibinfo{number}{24}) (\bibinfo{year}{1990})
  \bibinfo{pages}{9868--9872}, ISSN \bibinfo{issn}{0027-8424}.

\bibitem[{Duong et~al.(2002)Duong, Yacoub, Adriany, Hu, Ugurbil, Vaughan,
  Merkle, and Kim}]{Duong2002}
\bibinfo{author}{T.~Q. Duong}, \bibinfo{author}{E.~Yacoub},
  \bibinfo{author}{G.~Adriany}, \bibinfo{author}{X.~Hu},
  \bibinfo{author}{K.~Ugurbil}, \bibinfo{author}{J.~T. Vaughan},
  \bibinfo{author}{H.~Merkle}, \bibinfo{author}{S.-G. Kim},
  \bibinfo{title}{High-resolution, spin-echo {BOLD}, and CBF f{MRI} at 4 and 7
  {T}}, \bibinfo{journal}{Magn Reson Med}
  \bibinfo{volume}{48}~(\bibinfo{number}{4}) (\bibinfo{year}{2002})
  \bibinfo{pages}{589--593}, ISSN \bibinfo{issn}{1522-2594}.

\bibitem[{Yacoub et~al.(2003)Yacoub, Duong, Van De~Moortele, Lindquist,
  Adriany, Kim, UÄŸurbil, and Hu}]{Yacoub2003}
\bibinfo{author}{E.~Yacoub}, \bibinfo{author}{T.~Q. Duong},
  \bibinfo{author}{P.-F. Van De~Moortele}, \bibinfo{author}{M.~Lindquist},
  \bibinfo{author}{G.~Adriany}, \bibinfo{author}{S.-G. Kim},
  \bibinfo{author}{K.~UÄŸurbil}, \bibinfo{author}{X.~Hu},
  \bibinfo{title}{Spin-echo f{MRI} in humans using high spatial resolutions and
  high magnetic fields}, \bibinfo{journal}{Magn Reson Med}
  \bibinfo{volume}{49}~(\bibinfo{number}{4}) (\bibinfo{year}{2003})
  \bibinfo{pages}{655--664}, ISSN \bibinfo{issn}{1522-2594}.

\bibitem[{Harel(2012)}]{Harel2012}
\bibinfo{author}{N.~Harel}, \bibinfo{title}{Ultra high resolution f{MRI} at
  ultra-high field}, \bibinfo{journal}{NeuroImage}
  \bibinfo{volume}{62}~(\bibinfo{number}{2}) (\bibinfo{year}{2012})
  \bibinfo{pages}{1024--1028}, ISSN \bibinfo{issn}{1053-8119}.

\bibitem[{Yacoub et~al.(2008)Yacoub, Harel, and Uğurbil}]{Yacoub2008}
\bibinfo{author}{E.~Yacoub}, \bibinfo{author}{N.~Harel},
  \bibinfo{author}{K.~Uğurbil}, \bibinfo{title}{High-field f{MRI} unveils
  orientation columns in humans}, \bibinfo{journal}{Proc Natl Acad Sci USA}
  \bibinfo{volume}{105}~(\bibinfo{number}{30}) (\bibinfo{year}{2008})
  \bibinfo{pages}{10607--10612}.

\bibitem[{Logothetis et~al.(2002)Logothetis, Merkle, Augath, Trinath, and
  Ugurbil}]{Logothetis2002}
\bibinfo{author}{N.~K. Logothetis}, \bibinfo{author}{H.~Merkle},
  \bibinfo{author}{M.~Augath}, \bibinfo{author}{T.~Trinath},
  \bibinfo{author}{K.~Ugurbil}, \bibinfo{title}{Ultra High-Resolution f{MRI} in
  Monkeys with Implanted {RF} Coils}, \bibinfo{journal}{Neuron}
  \bibinfo{volume}{35}~(\bibinfo{number}{2}) (\bibinfo{year}{2002})
  \bibinfo{pages}{227--242}, ISSN \bibinfo{issn}{0896-6273}.

\bibitem[{Pfeuffer et~al.(2002)Pfeuffer, van~de Moortele, Yacoub, Shmuel,
  Adriany, Andersen, Merkle, Garwood, Ugurbil, and Hu}]{Pfeuffer2002}
\bibinfo{author}{J.~Pfeuffer}, \bibinfo{author}{P.-F. van~de Moortele},
  \bibinfo{author}{E.~Yacoub}, \bibinfo{author}{A.~Shmuel},
  \bibinfo{author}{G.~Adriany}, \bibinfo{author}{P.~Andersen},
  \bibinfo{author}{H.~Merkle}, \bibinfo{author}{M.~Garwood},
  \bibinfo{author}{K.~Ugurbil}, \bibinfo{author}{X.~Hu}, \bibinfo{title}{Zoomed
  Functional Imaging in the Human Brain at 7 Tesla with Simultaneous High
  Spatial and High Temporal Resolution}, \bibinfo{journal}{NeuroImage}
  \bibinfo{volume}{17}~(\bibinfo{number}{1}) (\bibinfo{year}{2002})
  \bibinfo{pages}{272--286}, ISSN \bibinfo{issn}{1053-8119}.

\bibitem[{Wu et~al.(2013)Wu, Tsai, Wu, Chuang, Shih, Chung, and Huang}]{Wu2013}
\bibinfo{author}{P.-H. Wu}, \bibinfo{author}{P.-H. Tsai},
  \bibinfo{author}{M.-L. Wu}, \bibinfo{author}{T.-C. Chuang},
  \bibinfo{author}{Y.-Y. Shih}, \bibinfo{author}{H.-W. Chung},
  \bibinfo{author}{T.-Y. Huang}, \bibinfo{title}{High spatial resolution brain
  functional MRI using submillimeter balanced steady-state free precession
  acquisitiona)}, \bibinfo{journal}{Med Phys}
  \bibinfo{volume}{40}~(\bibinfo{number}{12}) (\bibinfo{year}{2013})
  \bibinfo{pages}{122304--122316}.

\bibitem[{Hugger et~al.(2011)Hugger, Zahneisen, LeVan, Lee, Lee, Zaitsev, and
  Hennig}]{Hugger2011}
\bibinfo{author}{T.~Hugger}, \bibinfo{author}{B.~Zahneisen},
  \bibinfo{author}{P.~LeVan}, \bibinfo{author}{K.~J. Lee},
  \bibinfo{author}{H.-L. Lee}, \bibinfo{author}{M.~Zaitsev},
  \bibinfo{author}{J.~Hennig}, \bibinfo{title}{Fast Undersampled Functional
  Magnetic Resonance Imaging Using Nonlinear Regularized Parallel Image
  Reconstruction}, \bibinfo{journal}{PLoS One}
  \bibinfo{volume}{6}~(\bibinfo{number}{12}) (\bibinfo{year}{2011})
  \bibinfo{pages}{1--9}.

\bibitem[{Chaari et~al.(2014)Chaari, Ciuciu, M{\'e}riaux, and
  Pesquet}]{Chaari2014}
\bibinfo{author}{L.~Chaari}, \bibinfo{author}{P.~Ciuciu},
  \bibinfo{author}{S.~M{\'e}riaux}, \bibinfo{author}{J.-C. Pesquet},
  \bibinfo{title}{Spatio-temporal wavelet regularization for parallel {MRI}
  reconstruction: application to functional {MRI}}, \bibinfo{journal}{Magn
  Reson Mater Phy} \bibinfo{volume}{27}~(\bibinfo{number}{6})
  (\bibinfo{year}{2014}) \bibinfo{pages}{509--529}, ISSN
  \bibinfo{issn}{1352-8661}.

\bibitem[{Lu et~al.(2011)Lu, Li, Atkinson, and Vaswani}]{modCS}
\bibinfo{author}{W.~Lu}, \bibinfo{author}{T.~Li},
  \bibinfo{author}{I.~Atkinson}, \bibinfo{author}{N.~Vaswani},
  \bibinfo{title}{Modified-CS-residual for recursive reconstruction of highly
  undersampled functional {MRI} sequences}, in: \bibinfo{booktitle}{Proc Int
  Conf Image Process}, ISSN \bibinfo{issn}{1522-4880},
  \bibinfo{pages}{2689--2692}, \bibinfo{year}{2011}.

\bibitem[{Yan et~al.(2014)Yan, Nie, Wu, and Guo}]{Yan2014}
\bibinfo{author}{S.~Yan}, \bibinfo{author}{L.~Nie}, \bibinfo{author}{C.~Wu},
  \bibinfo{author}{Y.~Guo}, \bibinfo{title}{Linear Dynamic Sparse Modelling for
  functional {MR} imaging}, \bibinfo{journal}{Brain Inform}
  \bibinfo{volume}{1}~(\bibinfo{number}{1}) (\bibinfo{year}{2014})
  \bibinfo{pages}{11--18}, ISSN \bibinfo{issn}{2198-4018}.

\bibitem[{Han et~al.(2015)Han, Park, Kim, and Ye}]{han2015}
\bibinfo{author}{P.~K. Han}, \bibinfo{author}{S.-H. Park},
  \bibinfo{author}{S.-G. Kim}, \bibinfo{author}{J.~C. Ye},
  \bibinfo{title}{Compressed sensing for f{MRI} : feasibility study on the
  acceleration of non-{EPI} f{MRI} at 9.4 T}, \bibinfo{journal}{Biomed Res Int}
  \bibinfo{volume}{2015}~(\bibinfo{number}{131926}) (\bibinfo{year}{2015})
  \bibinfo{pages}{1--24}.

\bibitem[{Jung and Ye(2009)}]{Jung2009}
\bibinfo{author}{H.~Jung}, \bibinfo{author}{J.~C. Ye},
  \bibinfo{title}{Performance evaluation of accelerated functional {MRI}
  acquisition using compressed sensing}, in: \bibinfo{booktitle}{Proc IEEE Int
  Symp Biomed Imaging}, ISSN \bibinfo{issn}{1945-7928},
  \bibinfo{pages}{702--705}, \bibinfo{year}{2009}.

\bibitem[{Zong et~al.(2014)Zong, Lee, Poplawsky, Kim, and Ye}]{Zong2014}
\bibinfo{author}{X.~Zong}, \bibinfo{author}{J.~Lee}, \bibinfo{author}{A.~J.
  Poplawsky}, \bibinfo{author}{S.-G. Kim}, \bibinfo{author}{J.~C. Ye},
  \bibinfo{title}{Compressed sensing f{MRI} using gradient-recalled echo and
  {EPI} sequences}, \bibinfo{journal}{NeuroImage} \bibinfo{volume}{92}
  (\bibinfo{year}{2014}) \bibinfo{pages}{312--321}, ISSN
  \bibinfo{issn}{1053-8119}.

\bibitem[{Chavarr{\'\i}as et~al.(2015)Chavarr{\'\i}as, Abascal, Montesinos, and
  Desco}]{chavarrias2015}
\bibinfo{author}{C.~Chavarr{\'\i}as}, \bibinfo{author}{J.~Abascal},
  \bibinfo{author}{P.~Montesinos}, \bibinfo{author}{M.~Desco},
  \bibinfo{title}{Exploitation of temporal redundancy in compressed sensing
  reconstruction of f{MRI} studies with a prior-based algorithm ({PICCS})},
  \bibinfo{journal}{Med phys} \bibinfo{volume}{42}~(\bibinfo{number}{7})
  (\bibinfo{year}{2015}) \bibinfo{pages}{3814--3821}.

\bibitem[{Holland et~al.(2013)Holland, Liu, Song, Mazerolle, Stevens, Sederman,
  Gladden, D'Arcy, Bowen, and Beyea}]{Holland2013}
\bibinfo{author}{D.~J. Holland}, \bibinfo{author}{C.~Liu},
  \bibinfo{author}{X.~Song}, \bibinfo{author}{E.~L. Mazerolle},
  \bibinfo{author}{M.~T. Stevens}, \bibinfo{author}{A.~J. Sederman},
  \bibinfo{author}{L.~F. Gladden}, \bibinfo{author}{R.~C.~N. D'Arcy},
  \bibinfo{author}{C.~V. Bowen}, \bibinfo{author}{S.~D. Beyea},
  \bibinfo{title}{Compressed sensing reconstruction improves sensitivity of
  variable density spiral f{MRI}}, \bibinfo{journal}{Magn Reson Med}
  \bibinfo{volume}{70}~(\bibinfo{number}{6}) (\bibinfo{year}{2013})
  \bibinfo{pages}{1634--1643}, ISSN \bibinfo{issn}{1522-2594}.

\bibitem[{Fang et~al.(2016)Fang, Van~Le, Choy, and Lee}]{Fang2015}
\bibinfo{author}{Z.~Fang}, \bibinfo{author}{N.~Van~Le},
  \bibinfo{author}{M.~Choy}, \bibinfo{author}{J.~H. Lee}, \bibinfo{title}{High
  spatial resolution compressed sensing ({HSPARSE}) functional {MRI}},
  \bibinfo{journal}{Magn Reson Med} \bibinfo{volume}{76}~(\bibinfo{number}{2})
  (\bibinfo{year}{2016}) \bibinfo{pages}{440--455}, ISSN
  \bibinfo{issn}{1522-2594}.

\bibitem[{Chiew et~al.(2016)Chiew, Graedel, McNab, Smith, and
  Miller}]{Chiew2016}
\bibinfo{author}{M.~Chiew}, \bibinfo{author}{N.~N. Graedel},
  \bibinfo{author}{J.~A. McNab}, \bibinfo{author}{S.~M. Smith},
  \bibinfo{author}{K.~L. Miller}, \bibinfo{title}{Accelerating functional {MRI}
  using fixed-rank approximations and radial-cartesian sampling},
  \bibinfo{journal}{Magn Reson Med} \bibinfo{volume}{76}~(\bibinfo{number}{6})
  (\bibinfo{year}{2016}) \bibinfo{pages}{1825--1836}, ISSN
  \bibinfo{issn}{1522-2594}.

\bibitem[{Singh et~al.(2015)Singh, Tewfik, and Ress}]{Singh2015}
\bibinfo{author}{V.~Singh}, \bibinfo{author}{A.~Tewfik},
  \bibinfo{author}{D.~Ress}, \bibinfo{title}{Under-sampled functional {MRI}
  using low-rank plus sparse matrix decomposition}, in:
  \bibinfo{booktitle}{Proc IEEE Int Conf Acoust Speech Signal Process}, ISSN
  \bibinfo{issn}{1520-6149}, \bibinfo{pages}{897--901}, \bibinfo{year}{2015}.

\bibitem[{Donoho(2006)}]{Donoho2006}
\bibinfo{author}{D.~Donoho}, \bibinfo{title}{Compressed sensing},
  \bibinfo{journal}{IEEE Trans. Inf. Theory}
  \bibinfo{volume}{52}~(\bibinfo{number}{4}) (\bibinfo{year}{2006})
  \bibinfo{pages}{1289--1306}, ISSN \bibinfo{issn}{0018-9448}.

\bibitem[{Aggarwal et~al.(2017)Aggarwal, Shrivastava, Kabra, and
  Gupta}]{Aggarwal2017}
\bibinfo{author}{P.~Aggarwal}, \bibinfo{author}{P.~Shrivastava},
  \bibinfo{author}{T.~Kabra}, \bibinfo{author}{A.~Gupta},
  \bibinfo{title}{Optshrink LR+S: accelerated fMRI reconstruction using
  non-convex optimal singular value shrinkage}, \bibinfo{journal}{Brain Inform}
  \bibinfo{volume}{4}~(\bibinfo{number}{1}) (\bibinfo{year}{2017})
  \bibinfo{pages}{65--83}, ISSN \bibinfo{issn}{2198-4026}.

\bibitem[{Aggarwal and Gupta(2016)}]{Aggarwal2016}
\bibinfo{author}{P.~Aggarwal}, \bibinfo{author}{A.~Gupta},
  \bibinfo{title}{Accelerated f{MRI} reconstruction using Matrix Completion
  with Sparse Recovery via Split Bregman}, \bibinfo{journal}{Neurocomputing}
  \bibinfo{volume}{216} (\bibinfo{year}{2016}) \bibinfo{pages}{319--330}, ISSN
  \bibinfo{issn}{0925-2312}.

\bibitem[{Candes et~al.(2006)Candes, Romberg, and Tao}]{Candes2006}
\bibinfo{author}{E.~Candes}, \bibinfo{author}{J.~Romberg},
  \bibinfo{author}{T.~Tao}, \bibinfo{title}{Robust uncertainty principles:
  exact signal reconstruction from highly incomplete frequency information},
  \bibinfo{journal}{IEEE Trans. Inf. Theory}
  \bibinfo{volume}{52}~(\bibinfo{number}{2}) (\bibinfo{year}{2006})
  \bibinfo{pages}{489--509}, ISSN \bibinfo{issn}{0018-9448}.

\bibitem[{Wang et~al.(2008)Wang, Yang, Yin, and Zhang}]{wang2008}
\bibinfo{author}{Y.~Wang}, \bibinfo{author}{J.~Yang}, \bibinfo{author}{W.~Yin},
  \bibinfo{author}{Y.~Zhang}, \bibinfo{title}{A new alternating minimization
  algorithm for total variation image reconstruction}, \bibinfo{journal}{SIAM J
  Imaging Sci} \bibinfo{volume}{1}~(\bibinfo{number}{3}) (\bibinfo{year}{2008})
  \bibinfo{pages}{248--272}.

\bibitem[{Biswal et~al.(2010)Biswal, Mennes, Zuo, Gohel, Kelly, Smith,
  Beckmann, Adelstein, Buckner, Colcombe et~al.}]{biswal2010}
\bibinfo{author}{B.~B. Biswal}, \bibinfo{author}{M.~Mennes},
  \bibinfo{author}{X.-N. Zuo}, \bibinfo{author}{S.~Gohel},
  \bibinfo{author}{C.~Kelly}, \bibinfo{author}{S.~M. Smith},
  \bibinfo{author}{C.~F. Beckmann}, \bibinfo{author}{J.~S. Adelstein},
  \bibinfo{author}{R.~L. Buckner}, \bibinfo{author}{S.~Colcombe}, et~al.,
  \bibinfo{title}{Toward discovery science of human brain function},
  \bibinfo{journal}{Proc Natl Acad Sci USA}
  \bibinfo{volume}{107}~(\bibinfo{number}{10}) (\bibinfo{year}{2010})
  \bibinfo{pages}{4734--4739}.

\bibitem[{Boyd et~al.(2011)Boyd, Parikh, Chu, Peleato, and Eckstein}]{boyd2011}
\bibinfo{author}{S.~Boyd}, \bibinfo{author}{N.~Parikh},
  \bibinfo{author}{E.~Chu}, \bibinfo{author}{B.~Peleato},
  \bibinfo{author}{J.~Eckstein}, \bibinfo{title}{Distributed optimization and
  statistical learning via the alternating direction method of multipliers},
  \bibinfo{journal}{Found Trends Mach Learn}
  \bibinfo{volume}{3}~(\bibinfo{number}{1}) (\bibinfo{year}{2011})
  \bibinfo{pages}{1--122}.

\bibitem[{Chen et~al.(2016)Chen, O'Sullivan, Politte, Evans, Han, Whiting, and
  Williamson}]{Chen2016}
\bibinfo{author}{Y.~Chen}, \bibinfo{author}{J.~A. O'Sullivan},
  \bibinfo{author}{D.~G. Politte}, \bibinfo{author}{J.~D. Evans},
  \bibinfo{author}{D.~Han}, \bibinfo{author}{B.~R. Whiting},
  \bibinfo{author}{J.~F. Williamson}, \bibinfo{title}{Line Integral Alternating
  Minimization Algorithm for Dual-Energy X-Ray {CT} Image Reconstruction},
  \bibinfo{journal}{IEEE Trans Med Imaging}
  \bibinfo{volume}{35}~(\bibinfo{number}{2}) (\bibinfo{year}{2016})
  \bibinfo{pages}{685--698}, ISSN \bibinfo{issn}{0278-0062}.

\bibitem[{Monti et~al.(2014)Monti, Hellyer, Sharp, Leech, Anagnostopoulos, and
  Montana}]{Monti2014}
\bibinfo{author}{R.~P. Monti}, \bibinfo{author}{P.~Hellyer},
  \bibinfo{author}{D.~Sharp}, \bibinfo{author}{R.~Leech},
  \bibinfo{author}{C.~Anagnostopoulos}, \bibinfo{author}{G.~Montana},
  \bibinfo{title}{Estimating time-varying brain connectivity networks from
  functional {MRI} time series}, \bibinfo{journal}{NeuroImage}
  \bibinfo{volume}{103} (\bibinfo{year}{2014}) \bibinfo{pages}{427--443}, ISSN
  \bibinfo{issn}{1053-8119}.

\bibitem[{Zheng et~al.(2013)Zheng, Zhang, Yang, and Jiao}]{Zheng2013}
\bibinfo{author}{Y.~Zheng}, \bibinfo{author}{X.~Zhang},
  \bibinfo{author}{S.~Yang}, \bibinfo{author}{L.~Jiao},
  \bibinfo{title}{Low-rank representation with local constraint for graph
  construction}, \bibinfo{journal}{Neurocomputing} \bibinfo{volume}{122}
  (\bibinfo{year}{2013}) \bibinfo{pages}{398--405}, ISSN
  \bibinfo{issn}{0925-2312}.

\bibitem[{Lingala et~al.(2011)Lingala, Hu, DiBella, and Jacob}]{lingala2011}
\bibinfo{author}{S.~G. Lingala}, \bibinfo{author}{Y.~Hu},
  \bibinfo{author}{E.~DiBella}, \bibinfo{author}{M.~Jacob},
  \bibinfo{title}{Accelerated dynamic {MRI} exploiting sparsity and low-rank
  structure: kt {SLR}}, \bibinfo{journal}{IEEE Trans Med Imaging}
  \bibinfo{volume}{30}~(\bibinfo{number}{5}) (\bibinfo{year}{2011})
  \bibinfo{pages}{1042--1054}.

\bibitem[{Zhang et~al.(2010)Zhang, Block, and Frahm}]{radial}
\bibinfo{author}{S.~Zhang}, \bibinfo{author}{K.~T. Block},
  \bibinfo{author}{J.~Frahm}, \bibinfo{title}{Magnetic resonance imaging in
  real time: Advances using radial {FLASH}}, \bibinfo{journal}{J Magn Reson
  Imaging} \bibinfo{volume}{31}~(\bibinfo{number}{1}) (\bibinfo{year}{2010})
  \bibinfo{pages}{101--109}, ISSN \bibinfo{issn}{1522-2586}.

\bibitem[{Hansen(1992)}]{Lcurve}
\bibinfo{author}{P.~C. Hansen}, \bibinfo{title}{Analysis of Discrete Ill-posed
  Problems by Means of the L-curve}, \bibinfo{journal}{SIAM Rev}
  \bibinfo{volume}{34}~(\bibinfo{number}{4}) (\bibinfo{year}{1992})
  \bibinfo{pages}{561--580}, ISSN \bibinfo{issn}{0036-1445}.

\bibitem[{Lustig et~al.(2007)Lustig, Donoho, and Pauly}]{lustig2007sparse}
\bibinfo{author}{M.~Lustig}, \bibinfo{author}{D.~Donoho},
  \bibinfo{author}{J.~M. Pauly}, \bibinfo{title}{Sparse {MRI}: The application
  of compressed sensing for rapid MR imaging}, \bibinfo{journal}{Magn Reson
  Med} \bibinfo{volume}{58}~(\bibinfo{number}{6}) (\bibinfo{year}{2007})
  \bibinfo{pages}{1182--1195}.

\bibitem[{Wang and Bovik(2009)}]{Wang2009}
\bibinfo{author}{Z.~Wang}, \bibinfo{author}{A.~Bovik}, \bibinfo{title}{Mean
  squared error: Love it or leave it? A new look at Signal Fidelity Measures},
  \bibinfo{journal}{IEEE Signal Process Mag}
  \bibinfo{volume}{26}~(\bibinfo{number}{1}) (\bibinfo{year}{2009})
  \bibinfo{pages}{98--117}, ISSN \bibinfo{issn}{1053-5888}.

\bibitem[{Calhoun et~al.(2013)Calhoun, Potluru, Phlypo, Silva, Pearlmutter,
  Caprihan, Plis, and Adal{\i}}]{calhoun2013}
\bibinfo{author}{V.~D. Calhoun}, \bibinfo{author}{V.~K. Potluru},
  \bibinfo{author}{R.~Phlypo}, \bibinfo{author}{R.~F. Silva},
  \bibinfo{author}{B.~A. Pearlmutter}, \bibinfo{author}{A.~Caprihan},
  \bibinfo{author}{S.~M. Plis}, \bibinfo{author}{T.~Adal{\i}},
  \bibinfo{title}{Independent component analysis for brain f{MRI} does indeed
  select for maximal independence}, \bibinfo{journal}{PLoS One}
  \bibinfo{volume}{8}~(\bibinfo{number}{8}) (\bibinfo{year}{2013})
  \bibinfo{pages}{e73309}.

\bibitem[{Allen et~al.(2014)Allen, Damaraju, Plis, Erhardt, Eichele, and
  Calhoun}]{Allen2012}
\bibinfo{author}{E.~A. Allen}, \bibinfo{author}{E.~Damaraju},
  \bibinfo{author}{S.~M. Plis}, \bibinfo{author}{E.~B. Erhardt},
  \bibinfo{author}{T.~Eichele}, \bibinfo{author}{V.~D. Calhoun},
  \bibinfo{title}{Tracking Whole-Brain Connectivity Dynamics in the Resting
  State}, \bibinfo{journal}{Cereb Cortex}
  \bibinfo{volume}{24}~(\bibinfo{number}{3}) (\bibinfo{year}{2014})
  \bibinfo{pages}{663--676}.

\bibitem[{Kiviniemi et~al.(2009)Kiviniemi, Starck, Remes, Long, Nikkinen,
  Haapea, Veijola, Moilanen, Isohanni, Zang et~al.}]{kiviniemi2009}
\bibinfo{author}{V.~Kiviniemi}, \bibinfo{author}{T.~Starck},
  \bibinfo{author}{J.~Remes}, \bibinfo{author}{X.~Long},
  \bibinfo{author}{J.~Nikkinen}, \bibinfo{author}{M.~Haapea},
  \bibinfo{author}{J.~Veijola}, \bibinfo{author}{I.~Moilanen},
  \bibinfo{author}{M.~Isohanni}, \bibinfo{author}{Y.-F. Zang}, et~al.,
  \bibinfo{title}{Functional segmentation of the brain cortex using high model
  order group { PICA}}, \bibinfo{journal}{Hum Brain Mapp}
  \bibinfo{volume}{30}~(\bibinfo{number}{12}) (\bibinfo{year}{2009})
  \bibinfo{pages}{3865--3886}.

\bibitem[{Abou-Elseoud et~al.(2010)Abou-Elseoud, Starck, Remes, Nikkinen,
  Tervonen, and Kiviniemi}]{abou2010}
\bibinfo{author}{A.~Abou-Elseoud}, \bibinfo{author}{T.~Starck},
  \bibinfo{author}{J.~Remes}, \bibinfo{author}{J.~Nikkinen},
  \bibinfo{author}{O.~Tervonen}, \bibinfo{author}{V.~Kiviniemi},
  \bibinfo{title}{The effect of model order selection in group {PICA}},
  \bibinfo{journal}{Hum Brain Mapp} \bibinfo{volume}{31}~(\bibinfo{number}{8})
  (\bibinfo{year}{2010}) \bibinfo{pages}{1207--1216}.

\bibitem[{Smith et~al.(2009)Smith, Fox, Miller, Glahn, Fox, Mackay, Filippini,
  Watkins, Toro, Laird et~al.}]{smith2009}
\bibinfo{author}{S.~M. Smith}, \bibinfo{author}{P.~T. Fox},
  \bibinfo{author}{K.~L. Miller}, \bibinfo{author}{D.~C. Glahn},
  \bibinfo{author}{P.~M. Fox}, \bibinfo{author}{C.~E. Mackay},
  \bibinfo{author}{N.~Filippini}, \bibinfo{author}{K.~E. Watkins},
  \bibinfo{author}{R.~Toro}, \bibinfo{author}{A.~R. Laird}, et~al.,
  \bibinfo{title}{Correspondence of the brain's functional architecture during
  activation and rest}, \bibinfo{journal}{Proc Natl Acad Sci USA}
  \bibinfo{volume}{106}~(\bibinfo{number}{31}) (\bibinfo{year}{2009})
  \bibinfo{pages}{13040--13045}.

\bibitem[{Bell and Sejnowski(1995)}]{bell1995}
\bibinfo{author}{A.~J. Bell}, \bibinfo{author}{T.~J. Sejnowski},
  \bibinfo{title}{An information-maximization approach to blind separation and
  blind deconvolution}, \bibinfo{journal}{Neural Comput}
  \bibinfo{volume}{7}~(\bibinfo{number}{6}) (\bibinfo{year}{1995})
  \bibinfo{pages}{1129--1159}.

\bibitem[{Esposito et~al.(2002)Esposito, Formisano, Seifritz, Goebel, Morrone,
  Tedeschi, and Di~Salle}]{esposito2002}
\bibinfo{author}{F.~Esposito}, \bibinfo{author}{E.~Formisano},
  \bibinfo{author}{E.~Seifritz}, \bibinfo{author}{R.~Goebel},
  \bibinfo{author}{R.~Morrone}, \bibinfo{author}{G.~Tedeschi},
  \bibinfo{author}{F.~Di~Salle}, \bibinfo{title}{Spatial independent component
  analysis of functional {MRI} time-series: To what extent do results depend on
  the algorithm used?}, \bibinfo{journal}{Hum Brain Mapp}
  \bibinfo{volume}{16}~(\bibinfo{number}{3}) (\bibinfo{year}{2002})
  \bibinfo{pages}{146--157}.

\bibitem[{Tsao et~al.(2003)Tsao, Boesiger, and Pruessmann}]{tsao2003}
\bibinfo{author}{J.~Tsao}, \bibinfo{author}{P.~Boesiger},
  \bibinfo{author}{K.~P. Pruessmann}, \bibinfo{title}{k-t {BLAST} and k-t
  {SENSE}: Dynamic {MRI} with high frame rate exploiting spatiotemporal
  correlations}, \bibinfo{journal}{Magn Reson Med}
  \bibinfo{volume}{50}~(\bibinfo{number}{5}) (\bibinfo{year}{2003})
  \bibinfo{pages}{1031--1042}.

\bibitem[{Tsao et~al.(2005)Tsao, Kozerke, Boesiger, and Pruessmann}]{tsao2005}
\bibinfo{author}{J.~Tsao}, \bibinfo{author}{S.~Kozerke},
  \bibinfo{author}{P.~Boesiger}, \bibinfo{author}{K.~P. Pruessmann},
  \bibinfo{title}{Optimizing spatiotemporal sampling for k-t {BLAST} and k-t
  {SENSE}: Application to high-resolution real-time cardiac steady-state free
  precession}, \bibinfo{journal}{Magn Reson Med}
  \bibinfo{volume}{53}~(\bibinfo{number}{6}) (\bibinfo{year}{2005})
  \bibinfo{pages}{1372--1382}.

\bibitem[{Huang et~al.(2005)Huang, Akao, Vijayakumar, Duensing, and
  Limkeman}]{huang2005}
\bibinfo{author}{F.~Huang}, \bibinfo{author}{J.~Akao},
  \bibinfo{author}{S.~Vijayakumar}, \bibinfo{author}{G.~R. Duensing},
  \bibinfo{author}{M.~Limkeman}, \bibinfo{title}{k-t {GRAPPA}: A k-space
  implementation for dynamic {MRI} with high reduction factor},
  \bibinfo{journal}{Magn Reson Med} \bibinfo{volume}{54}~(\bibinfo{number}{5})
  (\bibinfo{year}{2005}) \bibinfo{pages}{1172--1184}.

\bibitem[{Van~Reeth et~al.(2012)Van~Reeth, Tham, Tan, and Poh}]{Reeth2012}
\bibinfo{author}{E.~Van~Reeth}, \bibinfo{author}{I.~W.~K. Tham},
  \bibinfo{author}{C.~H. Tan}, \bibinfo{author}{C.~L. Poh},
  \bibinfo{title}{Super-resolution in magnetic resonance imaging: A review},
  \bibinfo{journal}{Concepts in Magnetic Resonance Part A}
  \bibinfo{volume}{40A}~(\bibinfo{number}{6}) (\bibinfo{year}{2012})
  \bibinfo{pages}{306--325}.

\bibitem[{Feinberg and Setsompop(2013)}]{Feinberg2013}
\bibinfo{author}{D.~A. Feinberg}, \bibinfo{author}{K.~Setsompop},
  \bibinfo{title}{Ultra-fast MRI of the human brain with simultaneous
  multi-slice imaging}, \bibinfo{journal}{Journal of Magnetic Resonance}
  \bibinfo{volume}{229} (\bibinfo{year}{2013}) \bibinfo{pages}{90 -- 100},
  \bibinfo{note}{frontiers of In Vivo and Materials MRI Research}.

\end{thebibliography}
\bibliographystyle{elsarticle-num-names}

\end{document}